\documentstyle[cite,11pt,epsf]{article}

\columnwidth\textwidth
	
\newcommand{\nn}{\nonumber}
\newcommand{\be}{\begin{equation}}
\newcommand{\ee}{\end{equation}}
\newcommand{\bea}{\begin{eqnarray}}
\newcommand{\eea}{\end{eqnarray}}
\newcommand{\ggpp}{\gamma\gamma \rightarrow \pi ^+\pi^-}
\newcommand{\ggppp}{\gamma\gamma \rightarrow \pi ^1\pi^1}
\newcommand{\ggppn}{\gamma\gamma \rightarrow \pi ^0\pi^0}
\newcommand{\gpgp}{\gamma\pi^+ \rightarrow \gamma \pi^+}
\newcommand{\gpmgpm}{\gamma\pi^\pm \rightarrow \gamma \pi^\pm}

\newcommand{\sss}{\scriptscriptstyle}

\newcommand{\dac}{\Delta_A^{\sss{C}}}
\newcommand{\dbc}{\Delta_B^{\sss{C}}}
\newcommand{\apmb}{(\alpha \pm \beta)^{\sss{C}}}
\newcommand{\apmbn}{(\alpha \pm \beta)^{\sss{N}}}
\newcommand{\apb}{(\alpha + \beta)^{\sss{C}}}
\newcommand{\amb}{(\alpha - \beta)^{\sss{C}}}
\newcommand{\la}{\;\langle\;}

\newcommand{\ra}{\;\rangle\;}

\renewcommand{\theequation}{\arabic{section}.\arabic{equation}}
\def\mytoday#1{{ } \ifcase\month \or
 January\or February\or March\or April\or May\or June\or
 July\or August\or September\or October\or November\or December\fi
 \space \number\year}

\begin{document}
\noindent
\begin{tabbing} \` BUTP--96/02\\ \end{tabbing}
\vspace*{1.5cm}

\begin{center}

{\Large{\bf{Charged Pion--Pair Production and Pion Polarizabilities to two Loops 
\footnote{Work supported in part by Schweizerischer Nationalfonds}}}\\[1.5cm]}

U. B\"urgi\\[1cm] 

Institut f\"ur Theoretische Physik \\
Universit\"at Bern\\
Sidlerstrasse 5, CH--3012 Bern, Switzerland \\ [0.5cm]
(e--mail: buergi@butp.unibe.ch) \\[1.5cm]

\mytoday \\
\vspace{1cm}

{\bf{Abstract}}
\end{center}
\begin{quote}
We evaluate the amplitude for  $\ggpp$  to two loops in the framework of
chiral perturbation theory.
 The three new
 coupling constants that enter the result  at this order in the low--energy
expansion are estimated via resonance saturation.
 We discuss  in addition the crossed channel processes $\gpmgpm$ -- in
particular the charged pion polarizabilities --  to the same accuracy.
The predictions are compared with  available experimental
information.  
\end{quote}
 
\thispagestyle{empty}
\newpage

\renewcommand{\arraystretch}{1.5}
\setcounter{page}{2}
\setcounter{section}{0}
\setcounter{equation}{0}
\setcounter{subsection}{0}

\section{Introduction}

Pion--pair production by photons at low energies may be used to test chiral
perturbation theory (CHPT) \cite{WE79,GL84,GL85,LE94,chiral}. For neutral
pions, the chiral calculation at leading order -- generated by one--loop
diagrams \cite{BC88,DO88} -- is not in good agreement with the available data
from Crystal Ball
\cite{crystal} even near the threshold. Recently, the next--to--leading order
contribution, generated by two--loop diagrams, has been determined in Ref.
 \cite{BE94}.  The   corresponding enhancement of the
cross section  brings theory and experiment into
agreement within the uncertainties of  the calculation and the available
data \cite{BE94}.

In the process $\ggpp$,
the Born term (that is absent in the neutral channel)
dominates
the behavior of the  cross section at low energies. Next--to--leading
corrections (one--loop graphs) change the result very little \cite{BC88}, and
the comparison with the available data from Mark II \cite{mark}  works quite
well.
The dispersive analysis performed in \cite{DH93,PE95} indicates how additional
(multiloop) effects required by unitarity may influence the amplitude.
 The authors find only a slight change as compared to the one--loop
 result for center--of--mass energies $\sqrt{s} \le 700$ MeV. It is one of the
purposes of the present article to investigate
the effect of the two--loop contributions to the amplitude.

The Compton amplitude $\gpmgpm$  may be obtained from
 $\ggpp$ by  crossing.  The leading term in the Taylor
expansion of the Compton amplitude at threshold in powers of photon energies
measures the charge of the pions, whereas the next--to--leading terms are
parametrized through the electric  ($\bar{\alpha}_{\pi}$) and
magnetic ($\bar{\beta}_{\pi}$)  polarizabilities. These probe the
rigidity of the charged pion  against an external
electromagnetic perturbation.
 At leading order
in the chiral expansion, the result is \cite{TE73,DH89,HO90}
$\bar{\alpha}_{\pi}=-\bar{\beta}_{\pi} = 2.7 \pm 0.4$ \footnote{We express the
polarizabilities in units of $10^{-4}$ fm$^3$ throughout.}.
 With the two--loop Compton amplitude at hand, it is
then straightforward to evaluate
$\bar{\alpha}_{\pi}$ and
$\bar{\beta}_{\pi}$
  at next--to--leading order
in the quark mass expansion. The result of this analysis is
presented below. Turning  to the experimental
aspect, we note that the   experimental determination of the charged pion
polarizabilities is not an easy task.
At Lebedev \cite{AI86}
and at Serpukhov
\cite{AN84},   relevant experiments have been performed some time ago.
Unfortunately the results of these measurements do not agree very
well among each other, such that a comparison with the theoretical prediction
is not conclusive. In order to clarify the situation, additional
experiments are planned \cite{BB95}.

The article is organized as follows:
 After setting up the kinematics  and illuminating some aspects
of CHPT, the effective lagrangians ${\cal L}_2, {\cal L}_4$ and ${\cal L}_6$
are discussed. For completeness we  display in section \ref{section5} the tree--
and one--loop result worked out earlier
in \cite{BC88}. We then evaluate all diagrams
occurring at ${\cal O}(p^6)$, including also the pion mass and the wave
function renormalization constant at two--loop order (sections
\ref{eval2loop}, \ref{section8}). The amplitude for $\ggpp$ is displayed
in section \ref{ampsec}, that also contains a numerical discussion
of the corresponding cross section. Section \ref{compsec}
 is devoted a discussion of Compton scattering and pion
electric and magnetic polarizabilities at two--loop order.
 Finally, a summary and concluding remarks are given
in section \ref{sumsec}. The   notation used in the text is collected in appendix
\ref{loopfunc}.

\setcounter{equation}{0} \setcounter{subsection}{0}

\section{Kinematics \label{kinsec}}

The amplitude describing the process $\ggpp$ may be extracted from the matrix
element
\be
\la \pi^+(p_1),\pi^-(p_2)
\;out\;|\;\gamma(q_1,\lambda_1),\gamma(q_2,\lambda_2) \; in \ra = i
(2\pi)^4\delta^4(P_f-P_i) \;T^{\ggpp},
\label{kin1}
\ee
where
\bea
T^{\ggpp} & = & e^2
\epsilon_1^{\mu}(q_1,\lambda_1)\epsilon_2^{\nu}(q_2,\lambda_2)
W_{\mu\nu}^{\ggpp} \; , \nn \\ W_{\mu\nu}^{\ggpp} & = & i \int \! dx e^{-i
(q_1 x+q_2 y)} \la\pi^+(p_1)\pi^-(p_2) \;out \;| \;T\; j_{\mu}(x) j_{\nu} (y) \;
| \; 0\ra.
\label{kin2}
\eea
$j^{\mu}$ denotes the electromagnetic current and $\alpha=e^2/4\pi \simeq
1/137$ is the  electromagnetic coupling.  It is convenient to change the pion
coordinates according to $(\pi^{\pm},\pi^0) \rightarrow (\pi^1,\pi^2,\pi^3)$ and
instead of  $\pi^+\pi^-$--production, we consider in the following the process
$\gamma\gamma\rightarrow \pi^1\pi^1$, with
\be
W_{\mu\nu}^{\ggpp} = -W_{\mu\nu}^{\ggppp} \doteq -V^{\sss{C}}_{\mu\nu} \; ,
\label{kin3}
\ee
where the relative minus sign stems from the Condon--Shortly phase convention.
The correlator
$V_{\mu\nu}$ may be decomposed into four
 Lorentz invariant amplitudes
$A^{\sss{C}},B^{\sss{C}},C^{\sss{C}},D^{\sss{C}}$ (see e.g. \cite{BE94}),
\bea
V^{\sss{C}}_{\mu\nu} & = & A^{\sss{C}}(s,t,u)\, T_{1\mu\nu} +
B^{\sss{C}}(s,t,u) \,T_{2\mu\nu} + C^{\sss{C}}(s,t,u) \,T_{3\mu\nu} +
D^{\sss{C}}(s,t,u) \,T_{4\mu\nu} \nn \; , \\[0.1cm]
 T_{1\mu\nu} & = &
\frac{s}{2} g_{\mu\nu} - q_{1\nu} q_{2\mu} \; ,\nn \\
 T_{2\mu\nu} & = & 2 s
\Delta_{\mu}\Delta_{\nu} - \nu^2 g_{\mu\nu} -
2\nu \, (q_{1\nu}\Delta_{\mu}-q_{2\mu}\Delta_{\nu}) \nn \; ,\\[0.2cm]
 T_{3\mu\nu}
& = & q_{1\mu}q_{2\nu} \nn \; ,\\[0.2cm] 
T_{4\mu\nu} & = &
s \, (q_{1\mu}\Delta_{\nu}-q_{2\nu}\Delta_{\mu})-\nu \, (q_{1\mu}q_{1\nu}+q_{2\mu}q_{2\nu})
\nn \; , \\[0.2cm] \Delta_{\mu} & = & (p_1-p_2)_{\mu} \; ,
\label{kin4}
\eea
where
\bea
s & = & (q_1+q_2)^2 = 4 \vec{q} \;^2 \; , \nn \\ t & = & (p_1-q_1)^2 =
M_{\pi}^2-2 \vec{q}\;^2 (1-\beta(s) \cos \theta ) \; , \nn \\ u & = &
(p_2-q_1)^2 = M_{\pi}^2-2 \vec{q}\;^2 (1+\beta(s) \cos \theta ) \; , \nn \\
\nu & = & t-u \; ,
\label{kin5}
\eea
are the standard Mandelstam variables, 
$\beta(s)=(1-4 M_{\pi}^2/s)^{1/2}$ is the velocity of the
produced pions and  $\theta$ denotes the scattering angle,
$\vec{q}_1 .\vec{p}_1 = |\vec{q}_1| |\vec{p}_1| \cos \theta$ 
(all quantities in the center--of--mass system).  
The tensor $V^{\sss{C}}_{\mu\nu}$ satisfies furthermore the
Ward identities
\be
q_1^{\mu} V^{\sss{C}}_{\mu\nu} = q_2^{\nu} V^{\sss{C}}_{\mu\nu} = 0 \; .
\label{kin6}
\ee
The amplitudes $A^{\sss{C}}$ and $B^{\sss{C}}$ are analytic functions of the
variables $s$, $t$ and $u$, symmetric under crossing $(t,u)\rightarrow
(u,t)$. The remaining quantities $C^{\sss{C}}$ and $D^{\sss{C}}$ do not
contribute to the cross section (gauge invariance).

The differential cross section for unpolarized photons in the center--of--mass
system reads with the normalization $\la \vec{p}_1\, | \,\vec{p}_2 \ra = 2
(2\pi)^3 p_1^0\delta^3(\vec{p}_1-\vec{p}_2)$,
\bea
\frac{d\sigma}{d\Omega}^{\gamma\gamma \rightarrow \pi^+\pi^-} & = &
\frac{\alpha^2 s}{32}\beta(s) H^{\sss{C}}(s,t) \; , \nn \\ H^{\sss{C}}(s,t) &
= & |H^{\sss{C}}_{++}|^2+|H^{\sss{C}}_{+-}|^2 \; .
\label{kin7}
\eea
Here we have introduced the helicity amplitudes
$H^{\sss{C}}_{+\pm}$ corresponding to photon helicity differences $\lambda=
0,2$, respectively. They are defined in terms of the amplitudes $A^{\sss{C}}$
and $B^{\sss{C}}$,
\bea
H^{\sss{C}}_{++} & = & A^{\sss{C}} + 2(4M_{\pi}^2-s)B^{\sss{C}} \; , \nn \\
H^{\sss{C}}_{+-} & = & \frac{8(M_{\pi}^4-tu)}{s}B^{\sss{C}} \; ,
\label{kin8}
\eea
and have partial wave expansion involving even $J \ge \lambda$,
\bea
H^{\sss{C}}_{++} & = & \sum_{J=0,2,4 \ldots} h_{\sss{C},+}^J (s)
d_{\sss{00}}^J(\cos \theta) \; , \nn \\ H^{\sss{C}}_{+-} & = & \sum_{J=2,4,6
\ldots} h_{\sss{C},-}^J (s) d_{\sss{20}}^J(\cos \theta) \; .
\label{kin85}
\eea
In the text we also use the symbols
\bea
\bar{H}^{\sss{C}}_{++} & \doteq & H^{\sss{C}}_{++} - H^{\sss{C}}_{{\sss{B}} ++} \; , \nn \\
\bar{H}^{\sss{C}}_{+-} & \doteq & H^{\sss{C}}_{+-} - H^{\sss{C}}_{{\sss{B}} +-} \; , 
\label{kin86}
\eea
where $H^{\sss{C}}_{{\sss{B}} +\pm}$ denote the helicity amplitudes generated
by the Born contributions.

For comparison with experimental results, it is convenient to present also the
total cross section for the case where $|\cos\theta \, |$ is less than some fixed
value $Z$ ,
\be
\sigma(s,|\cos\theta \, |<Z) = \frac{\alpha^2 \pi}{8} \int_{t_-}^{t_+} \! dt \,
H^{\sss{C}}(s,t)
\label{kin9}
\ee
with
\be
t_{\pm} = M_{\pi}^2-2\vec{q}\;^2 (1 \mp \beta(s) Z)\; .
\label{kin10}
\ee

\setcounter{equation}{0} \setcounter{subsection}{0}

\section{Low energy expansion}
We consider QCD with two flavours $u$ and $d$ in the isospin limit
$m_u=m_d=\hat{m}$. Since the masses of these two quarks are small, the
deviation from chiral symmetry may be studied by treating the quark mass term
as a perturbation using the framework of an effective chiral lagrangian (CHPT) 
\cite{WE79,GL84,GL85,LE94,chiral}.  
This effective
lagrangian relies only on the solid assumptions of spontaneously broken
$SU(2)_L \times SU(2)_R$ chiral symmetry, Lorentz invariance and low momentum
transfer. It can be expanded in a series of terms with increasing numbers of
derivatives and powers of quark masses,
\be
{\cal L}_{\sss{eff}} = {\cal L}_2 + {\cal L}_4 + {\cal L}_6 + \ldots ,
\ee
where ${\cal L}_{(2n)}$ denotes a term of order $p^{2n}$. Each of these terms
contain a couple of monomials with couplings which are not fixed by symmetry
constraints, but they are fixed by the dynamics of the underlying theory
through the renormalization group invariant scale $\Lambda_{\sss{QCD}}$ and the heavy
quark masses. Up to now it has not been possible to reliably calculate them directly from the
QCD lagrangian. The best way to determine their values is by comparison with
experimental data.  Green functions of the theory are generated by the
vacuum--to--vacuum amplitude \cite{GL84,GL85}
\be
e^{iZ(v,a,s,p)} = \la 0_{out}\, | \, 0_{in} \ra_{v,a,s,p} = \int [dU] e^{i \int \! d^4 \! x {\cal L}
_{eff} (U,v,a,s,p)} \; ,
\ee
where the dynamical variables (pions) are contained in the matrix $U$. The
generating functional $Z$ admits an expansion in powers of the external
momenta and the quark masses
\be
Z(v,a,s,p) = Z_2 + Z_4 + Z_6 + \ldots ,
\ee
where $Z_n$ denotes a term of order $p^n$. The contribution of loops to the
generating functional is suppressed with respect to tree diagrams. At order
$p^2$ the contribution to the amplitude stems from ${\cal L}_2$ only via tree
diagrams.  At next--to--leading order one has to evaluate one--loop graphs
generated by ${\cal L}_2$ and adding the tree graphs from ${\cal L}_4$.
Diagrams at ${\cal O}(p^6)$ contribute via $Z_6$ from ${\cal L}_2+{\cal
L}_4+{\cal L}_6$ with tree, from ${\cal L}_2+{\cal L}_4$ with one--loop and
from ${\cal L}_2$ with two--loop diagrams. The one--loop diagrams at ${\cal
O}(p^6)$ contain one vertex from ${\cal L}_4$.  We write the corresponding
expansion of the amplitude as
\be
I = I_2 + I_4 + I_6 + \ldots , \;\;\;\; I = V^{\sss{C}}_{\mu\nu}, A^{\sss{C}},
B^{\sss{C}} ,
\ee
where it is understood that $Z_n$ generates $I_n$. For comprehensive reviews
on chiral perturbation theory we refer the reader to \cite{chiral}.

\setcounter{equation}{0} \setcounter{subsection}{0}

\section{The effective lagrangians ${\cal L}_2$, ${\cal L}_4$ and ${\cal L}_6$ \label{section4}}
At ${\cal O}(p^2)$ the most general effective lagrangian coincides with the
non--linear $\sigma$--model lagrangian ${\cal L}_2$ equipped with external
fields \cite{GL84},
\bea
{\cal L}_2 & = & \frac{1}{4} F^2 \la D_{\mu} U D^{\mu} U^{\dagger} +
\chi^{\dagger} U + \chi U^{\dagger} \ra \nn \\ D_{\mu} U & = & \partial_{\mu}
U - i (v_{\mu}+a_{\mu}) U + i U (v_{\mu}-a_{\mu}) , \nn \\[0.15cm] \chi & = &
2 B \, (s + i p) \; .
\label{la1}
\eea
The brackets $\la\ldots\ra$ denote a trace in flavour space and the unitary
$2\times 2$ matrix $U$ contains the pion fields,
\bea
U & = & \sigma+i\frac{\phi}{F} \;\;\; ; \;\;\; \sigma^2+\frac{\phi^2}{F^2} =
{\bf 1}\; ; \nn \\ \phi & = & \phi^i \tau^i = \left( \begin{array}{cc} \pi^0 &
\sqrt{2} \;\pi^+ \\ \sqrt{2} \;\pi^- & -\pi^0 \end{array} \right) .
\label{la2}
\eea
We count the Goldstone fields as ${\cal O}(p^0)$, a derivative
$\partial_{\mu}$, the vector and axial--vector currents $v_{\mu}, a_{\mu}$ as
${\cal O}(p)$ and the scalar and pseudoscalar currents $s, p$ as ${\cal
O}(p^2)$.  The lagrangian ${\cal L}_2$ contains two free parameters $F$ and
$B$, where $F$ is the pion decay constant in the chiral limit and $B$ is
related to the order parameter $\la 0 \,|\, \bar{q} q \,|\, 0 \ra$. In terms of physical
quantities we have
\bea
F_{\pi} & = & F ( 1 + {\cal O}(\hat{m}) ) \;, \nn \\ M_{\pi}^2 & =& M^2 ( 1 +
{\cal O}(\hat{m}) ) \;, \;\; M^2 = 2 B \hat{m} .
\label{la3}
\eea
The transformation properties under $G=SU(2)_L\times SU(2)_R \times U(1)$ of
the several fields are,
\bea
U & \rightarrow & g_R U g_L^{\dagger} \; , \nn \\ v_{\mu}+a_{\mu} &
\rightarrow & g_R (v_{\mu}+a_{\mu}) g_R^{\dagger}+i g_R \partial_{\mu}
g_R^{\dagger} \; , \nn \\ v_{\mu}-a_{\mu} & \rightarrow & g_L
(v_{\mu}-a_{\mu}) g_L^{\dagger}+i g_L \partial_{\mu} g_L^{\dagger} \; , \nn \\
s + i p & \rightarrow & g_R (s + i p) g_L^{\dagger} \; , \nn \\ g_{L,R} & = &
e^{i \phi} V_{L,R} \; , \nn \\ V_{L,R} & \in & SU(2) \;\; ; \;\;
\phi=\mbox{diag} \, (\phi_0,\phi_0) \;\; ; \;\; \phi \in {\bf R} \; .
\label{la4}
\eea
Since the charge is not a generator of $SU(2)$, we consider in the following
the case
\be
\la a_{\mu} \ra = 0 \;\;\; ; \; \;\; \la v_{\mu} \ra \not= 0 \; . \nn
\label{la5}
\ee
This condition is consistent with the transformation law (\ref{la4}). To
calculate $V^{\sss{C}}_{\mu\nu}$, we set
\be
s = \hat{m} {\bf 1} \;\;\; ; \;\;\; v_{\mu} = Q \bar{v}_{\mu} \;\;\; ; \;\;\;
p = \bar{p}^i \tau^i \;\;\; ; \;\;\; a_{\mu} = 0 \; ,
\label{la6}
\ee
where $Q=1/3$ diag$(2,-1)$ is the charge matrix.

The next--to--leading term $Z_4$ contains two types of contributions: one--loop
graphs generated by ${\cal L}_2$ and tree graphs involving one vertex from
${\cal L}_4$. The most general chiral invariant lagrangian ${\cal L}_4$ was
constructed by Gasser and Leutwyler \cite{GL84}:
\bea
{\cal L}_4 & = & {\cal L}^{(4)} +{\cal L}_{WZW} \; , \nn \\ {\cal L}^{(4)} & =
& \sum_{i=1}^7 \; l_i P_i + \ldots
\label{la7}
\eea
where \\[0.1cm]
\parbox{0.5cm}{} \hfill
\parbox{5.5cm}{\bea P_1 & = & \frac{1}{4} \la u^{\mu} u_{\mu} \ra ^2
\; , \nn \\ P_2 & = & \frac{1}{4} \la u_{\mu} u_{\nu} \ra\la u^{\mu} u^{\nu}
\ra \; , \nn \\ P_3 & = & \frac{1}{16} \la \chi_+ \ra^2 \; , \nn \\ P_4 & = &
\frac{i}{4} \la u_{\mu} \chi_-^{\mu} \ra \; , \nn
\label{la8a}
\eea}
\hfill
\parbox{5cm}{\bea P_5 & = & -\frac{1}{2} \la f_-^{\mu\nu} f_{ -
\mu\nu} \ra \; , \nn \\ P_6 & = & \frac{i}{4} \la f_+^{\mu\nu} [
u_{\mu},u_{\nu}] \ra \; , \nn \\ P_7 & = & -\frac{1}{16} \la \chi_- \ra^2 \;
\nn .
\label{la8b}
\eea}
\hfill
\parbox{1cm}{\bea & & \label{la8} \eea} \\ The following
notation was used:
\bea
u_{\mu} & = & i u^{\dagger} D_{\mu} U u^{\dagger} = -i u D_{\mu} U^{\dagger} u
= u_{\mu}^{\dagger} \; , \nn \\ \chi_{\pm} & = & u^{\dagger} \chi u^{\dagger}
\pm u \chi^{\dagger} u \; , \nn \\ \chi_-^{\mu} & = & u^{\dagger} D^{\mu} \chi
u^{\dagger} - u D^{\mu} \chi^{\dagger} u \; ; \nn \\ f^{\mu\nu}_{\pm} & = & u
F_L^{\mu\nu} u^{\dagger} \pm u^{\dagger} F_R^{\mu\nu} u \; ,
\label{la9}
\eea
with $u^2=U$. The field strength tensors $F_{L,R}^{\mu\nu}$ are related to the
nonabelian external fields $r_{\mu}=v_{\mu}+a_{\mu}$,
$l_{\mu}=v_{\mu}-a_{\mu}$ through
\bea
F_R^{\mu\nu} & = & \partial_{\mu} r_{\nu}-\partial_{\nu} r_{\mu}-i \,[
r_{\mu},r_{\nu} ] \; , \nn \\ F_L^{\mu\nu} & = & \partial_{\mu}
l_{\nu}-\partial_{\nu} l_{\mu}-i \,[ l_{\mu},l_{\nu} ] \; .
\label{la10}
\eea
The anomaly term ${\cal L}_{WZW}$ contributes to $V^{\mu\nu}_8$ \cite{AM92};
we therefore simply neglect these contributions. The ellipses in (\ref{la7})
denote polynomials in the external fields which are independent of the pion
variable. These terms do not contribute to S--matrix elements. \\ The
realization of $G$ on the fields in Eq. (\ref{la9}) is given by
\be
I \stackrel{G}{\rightarrow} h(\phi) I h^{\dagger}(\phi) \; ,
\label{la12}
\ee
where the non--linear realization $h(\phi)$ defines the action of $G$ on a
coset element $u(\phi)$ via
\be
u(\phi) \stackrel{G}{\rightarrow} g_R u(\phi) h^{\dagger}(\phi) = h(\phi)
u(\phi) g_L^{\dagger} \; .
\label{la11}
\ee
Since the one--loop graphs generated by ${\cal L}_2$ may be ultraviolet
divergent, the low--energy constants $l_i$ have to absorb these divergences
leading to an ultraviolet finite representation of any contribution at ${\cal
O}(p^4)$. Using dimensional regularization, the low--energy constants are
defined \cite{GL84} as
\bea
l_i & = & l_i^r + \gamma_i \lambda \;\; , \;\; i=1,\ldots,7 \; , \nn \\
\lambda & = & \frac{\mu^{2\omega}}{16
\pi^2} \left\{\frac{1}{2\omega}-\frac{1}{2}(\ln 4\pi+\Gamma' (1)+1) \right\} \; ,
\label{la13}
\eea
where $l_i^r$ are the coupling constants renormalized at the scale $\mu$. The
$\gamma_i$ are displayed in Ref. \cite{GL84}.   

Terms at ${\cal O}(p^6)$ generated by $Z_6$ contain two--loop diagrams with 
vertices from ${\cal L}_2$, one--loop graphs with one vertex from ${\cal L}_4$
and tree diagrams generated by ${\cal L}_6$.
The lagrangian ${\cal L}_6$ in the mesonic sector
was recently constructed by Fearing and Scherer \cite{FS94}.
 In the
case of $SU(3)\times SU(3)$, they find 111 (32) independent terms in the even
(odd) intrinsic parity sector. In the even parity sector relevant here, the new
low--energy constants that enter at this order
 have the form
\be B_i = B_i^r +B_i^d \;\;\;\; i=1,\ldots,111 \; ,
\label{la15}
\ee
 where the $B_i^d$ contain poles and
double poles at $d=4$. 

The amplitude
generated by tree diagrams involving one vertex
from ${\cal L}_6$ contributes only a polynomial to $\ggpp$, which has the same
structure as the divergent part in the two--loop amplitude. To illustrate this
fact, we note that the monomials \cite{FS94} 
\bea
\delta {\cal L}_6 & = & B_{30} \la\,[\,G^{\alpha\beta}\,]_+\,[\,D_{\mu}D^{\mu}
                              G_{\alpha\beta}\,]_+\ra \nn \\ & + & B_{31}
                              \la\,[\,G^{\alpha\beta}\,]_+\,
[\,D_{\alpha}D^{\gamma}
                              G_{\beta\gamma}\,]_+\ra \nn \\ & + &B_{50}
                              \la\,[\,\chi\,]_+\ra\la\,
[\,G^{\mu\nu}\,]_+[\,G_{\mu\nu}\,]_+\ra
\eea
generate the polynomials
\bea
\delta A^{\sss{C}}_6 & = & \frac{2}{9 F^4} \,[\, (4 B_{30}+9 B_{31} ) s- ( 8
B_{30} +112 B_{50} ) M^2 \,]\, \; , \nn \\ \delta B^{\sss{C}}_6 & = &
\frac{9}{2 F^4} B_{31} \; .
\label{la17}
\eea
One may therefore remove the divergences in the two--loop amplitude by simply
dropping the singular part. The finite pieces one is left with may be
estimated via resonance exchange (see below). In the following, we use the
notation
\bea
\delta A^{\sss{C}}_6 &=& \frac{a_1^{r,c} M^2 + a_2^{r,c} s}{(16\pi^2 F^2)^2}\;
,\nn \\ \delta B^{\sss{C}}_6 &=& \frac{b^{r,c}}{(16\pi^2 F^2)^2} \; ,
\label{ren13}
\eea
for the total contribution from ${\cal L}_6$.

\setcounter{equation}{0} \setcounter{subsection}{0}

\section{Results at ${\cal O}(p^2),{\cal O}(p^4)$ \label{section5}}
The process $\ggpp$  occurs already at tree level,
 in contrast to the situation in the neutral channel, where the leading term
starts at one--loop order. The amplitude at tree level
coincides with the one obtained from scalar electrodynamics,
\begin{figure}[t]
\unitlength1cm
\begin{picture}(2.5,1) \end{picture}
\epsfysize=3cm \epsffile{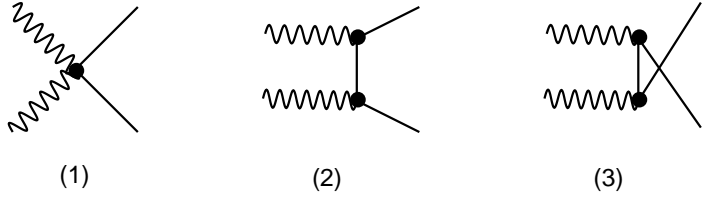}
\caption[Leading order Feynman diagrams]{\label{tree}\small{Tree diagrams
contributing to $Z_2$. The mass in the propagators is identified with the
physical pion mass at this order.}}
\end{figure}
\bea
A^{\sss{C}} & = & \hspace{0.5cm}\left[ \frac{1}{M^2-t}+\frac{1}{M^2-u} \right] + {\cal O}(1)
\; , \nn \\ B^{\sss{C}} & = & \frac{1}{2s} \left[
\frac{1}{M^2-t}+\frac{1}{M^2-u} \right] + {\cal O}(1) \; .
\label{amp1}
\eea
The pion mass in the propagators is identified with the physical one at this
order in the expansion. The corresponding Feynman diagrams are depicted in
Fig. \ref{tree}. The next--to--leading term $Z_4$ contains two types of
contributions: one--loop graphs generated by ${\cal L}_2$ and tree graphs
involving one vertex from ${\cal L}_4$. 
The term proportional to the low--energy constant $l_3$ in the lagrangian ${\cal L}_4$
is quadratic in the fields. One may therefore shift the pion mass squared according to
 $M^2 \rightarrow M_0^2=M^2+2 M^4 l_3/F^2$ and from now on we denote 
the pion propagator by $(M_0^2-p^2)^{-1}$, also in Fig. \ref{tree}.
Evaluating the several diagrams to
${\cal O}(p^4)$ (Fig. \ref{1loop}) gives \cite{BC88}
\bea
A^{\sss{C}} & = & \hspace{0.5cm} \left[ \frac{1}{M_{\pi}^2-t}+\frac{1}{M_{\pi}^2-u} \right]
+\frac{2}{F^2}\left[ \bar{G}(s)+\frac{\bar{l}_{\Delta}}{48\pi^2}\right] +
{\cal O}(p^2) \; , \nn \\ B^{\sss{C}} & = & \frac{1}{2s} \left[
\frac{1}{M_{\pi}^2-t}+\frac{1}{M_{\pi}^2-u} \right] + {\cal O}(1) \; .
\label{amp2}
\eea
The loop function
$\bar{G}(s)$ is discussed in appendix \ref{loopfunc}. The physical pion mass
squared $M_\pi^2$ is
\be
M_{\pi}^2 = M^2 \left[ 1+\frac{M^2}{F^2} \left(2
l_3^r+\frac{1}{32\pi^2}\ln\frac{M^2}{\mu^2}\right) + {\cal O}(M^4) \right] \;
.
\label{amp3}
\ee
The pion decay constant $F$ is identified at this order with the physical
$F_{\pi}$. It is a specific feature of the process $\gamma\gamma\rightarrow
\pi^+\pi^-$ that, after mass renormalization, the one--loop contribution from
 ${\cal L}_2$ is ultraviolet
finite. ${\cal L}_4$ generates the graphs (8,9) in Fig. \ref{1loop}. This
contribution is proportional to the finite, scale independent combination
\be
\bar{l}_{\Delta} = 96\pi^2(2l_5^r-l_6^r) = 2.7 \; .
\ee
 As is discussed in more detail in section \ref{crosssec},
there is good agreement with the available experimental data and the chiral
representation (\ref{amp2}).
We now turn to  the
evaluation of the contributions at order $p^6$.

\begin{figure}[t]
\unitlength1cm
\begin{picture}(2,1) \end{picture}
\epsfysize=5cm \epsffile{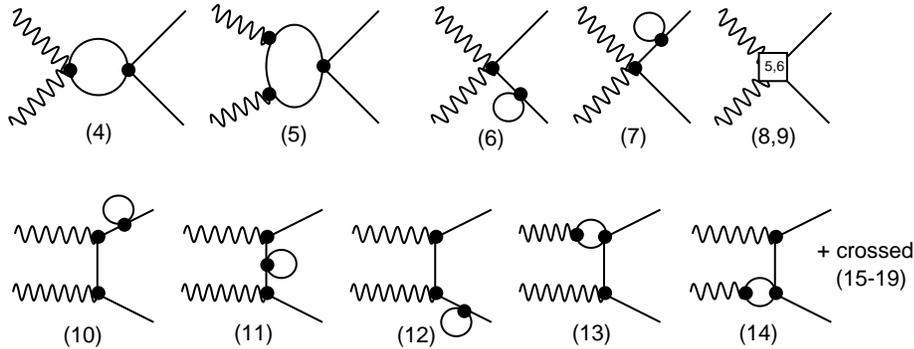}
\caption[Next--to--leading order Feynman diagrams]{\label{1loop}\small{One--loop
diagrams contributing to $Z_4$; the graphs (4, 5) generate the finite loop
function $\bar{G}(s)$. The graphs (10 --19) amounts to the replacement
$M^2 \rightarrow M_{\pi}^2$ in the tree result. Graphs (8, 9) stem from the
lagrangian ${\cal L}_4$.}}
\end{figure}

\setcounter{equation}{0} \setcounter{subsection}{0}

\section{Evaluation of diagrams at ${\cal O}(p^6)$ \label{eval2loop}}
\begin{figure}
\unitlength1cm
\begin{picture}(1,1) \end{picture}
\epsfysize=18cm \epsffile{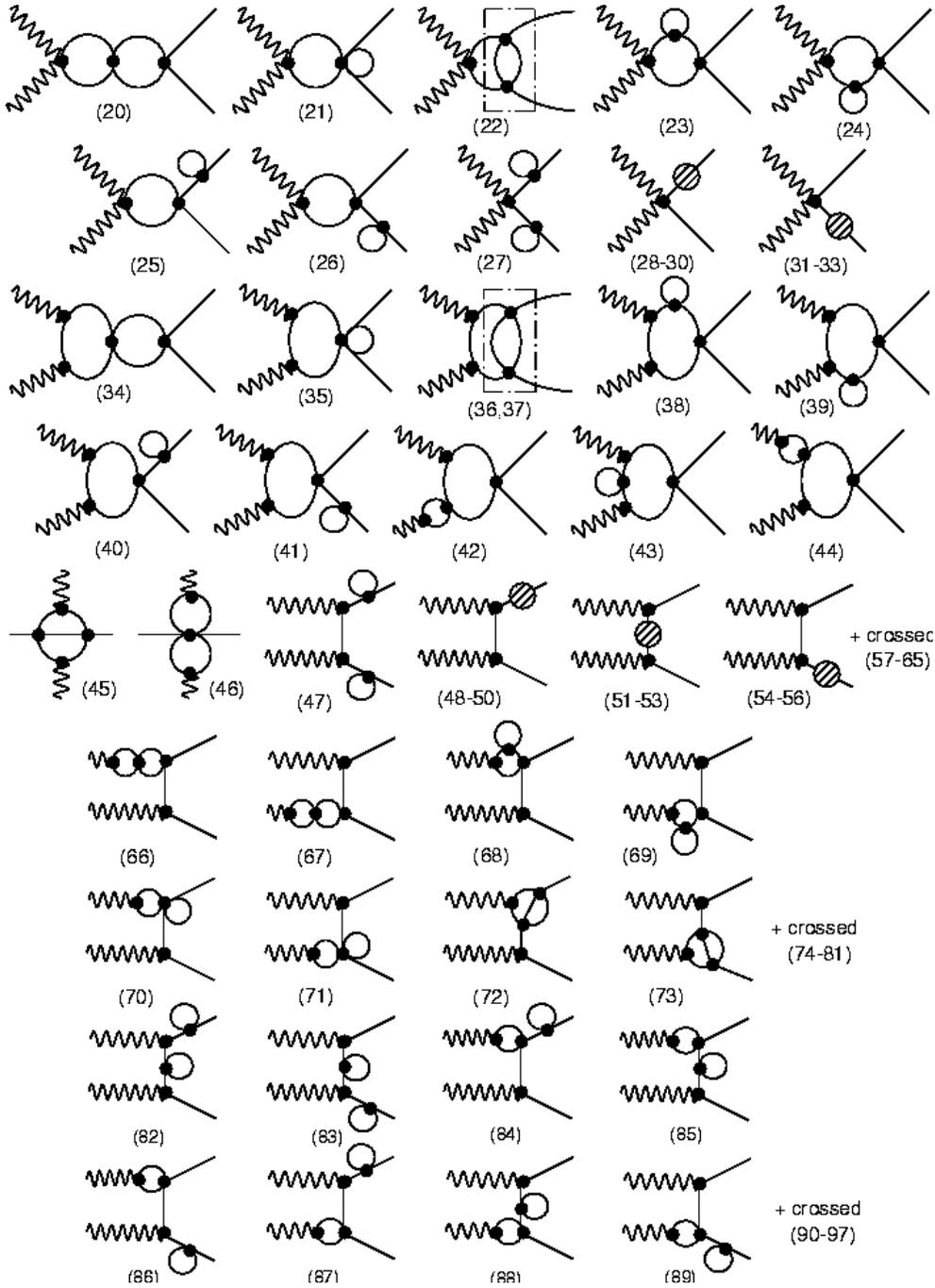}
\caption[Two--loop Feynman diagrams from ${\cal
L}_2$]{\label{2loopl2}\small{Full set of two--loop diagrams generated by ${\cal
L}_2$. The vertex (22), box (36,37) and acnode (45) graphs are genuine
two--loop diagrams, which do not have a representation in terms of one--loop
functions. The dash--dotted boxes indicate subdiagrams which we have treated in
the dispersive manner. We use symmetry arguments which relate the
contributions from the reducible diagrams (47--97) to mass-- and wavefunction
corrections at two--loop order. The hatched circles summarize selfenergy
contributions to the pion propagator (graphs (b -- d) in Fig. \ref{mass}). }}
\end{figure}

Up to ${\cal O}(p^4)$ one has to deal with only a few Feynman diagrams for the
process $\ggpp$ and the evaluation of the amplitude is straightforward. At
${\cal O}(p^6)$ we have found remarkably many diagrams. To generate all graphs and to check the
combinatorial factor for each of them, we use a beta--version of the {\it
Mathematica} package {\it FeynArts 2.0} developed by S. Kueblbeck and H. Eck
\cite{KE94}, where we have included the corresponding Feynman rules for CHPT,
adapted to $\ggpp$.  In Fig. \ref{2loopl2} we have plotted the full
set of two--loop diagrams generated by ${\cal L}_2$. We have not
distinguished the different intermediate pions running in the loops
$(\pi^1,\pi^2,\pi^3)$. Each displayed diagram is the generating one for a
class of subdiagrams with the same topology.

\subsection{Two--loop diagrams from ${\cal L}_2$ \label{secirr}}
\begin{figure}[tb]
\unitlength1cm
\begin{picture}(2,1) \end{picture}
\epsfysize=3cm \epsffile{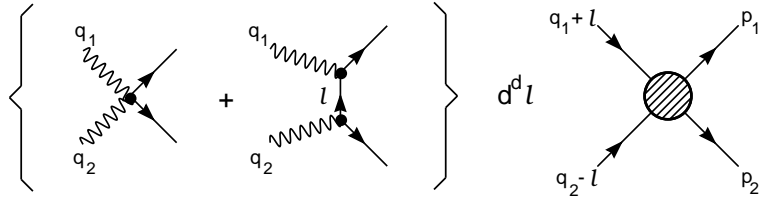}
\caption[Diagram generation with the $\pi\pi$ scattering
amplitude]{\label{pipi}\small{Construction scheme for a subclass of two--loop
diagrams. The four--point function on the right--hand side is the $d$--dimensional
elastic $\pi\pi$--amplitude at one--loop accuracy with two pions off--shell. The
particles running between left and right are $\pi^1$ and $\pi^2$. The symbol
$d^d l$ stands for integration over internal lines with weight (\ref{irr1}).}}
\end{figure}

The genuine two--loop diagrams 22 (vertex),  36,37
(box) and 45 (acnode) in Fig. \ref{2loopl2} cannot be represented as products
of one--loop integrals. A method to perform the relevant momentum integrals is
given in \cite{BE94,BU96a,BU96b}.
The graphs (20--26, 34--41) may be generated according to Fig. \ref{pipi}, where the four--point
function on the right--hand side is the $d$--dimensional elastic
$\pi\pi$--scattering amplitude at one--loop accuracy, with two pions
off--shell. The loop integration over internal momenta with weight
\be
\frac{1}{[M_{\pi}^2-(l+q_1)^2]\;[M_{\pi}^2-(q_2-l)^2]} \; ,
\label{irr1}
\ee
is indicated by the symbol $d^d l$ in the figure. $M_{\pi}^2$ denotes the
physical pion
mass in one--loop approximation (Eq. \ref{amp3}) and the momenta of the pions
running in the loop are $(l+q_1)^{\mu}$ and $(q_2-l)^{\mu}$. This procedure
has the advantage that the nonlocal singularities  generated by
divergent subdiagrams (enclosed
by dash--dotted lines in the graphs (22,36,37)) are automatically removed \cite{BE94}.
 
A further comment concerns diagram
(23) and (24). The tadpole in the loop propagator amounts to the replacement
$M^2 \rightarrow M^2_{\pi}$ if one also adds the one--loop diagram (4) in
Fig. \ref{1loop} multiplied with the wave function renormalization constant
$Z_{\pi}$ given at one--loop accuracy. A similar argument holds in the case of
diagrams (38,39,42--44). Summing up these diagrams allows one to introduce the
physical mass in the loop propagators.

The graphs (27--33) may be generated by
multiplying the seagull diagram (Fig. \ref{tree}, graph 3) with that part of the
wave function renormalization constant $Z_{\pi}$, which is generated by ${\cal L}_2$.
Furthermore, we use isospin symmetry arguments to
handle the diagrams (47--97) that also incorporate the pion propagator at
two--loop accuracy \cite{BU96a}.

\subsection{Pion propagator at ${\cal O}(p^6)$}
We compute the correlator
\be
\delta^{ab}\frac{1}{i}\Delta'_c(p^2)= \int\!d^D\! x e^{ipx}\la
0|T\phi^a(x)\phi^b(0)|0\ra
\ee
in the $\sigma$-- model parametrization Eq. (\ref{la2}). The complete propagator can be written
in terms of the bare
propagator$\Delta'(p^2)=\frac{1}{i}\frac{1}{M^2-p^2-i\epsilon}$ and the self
energy function $\Sigma(p^2)$ in the standard manner,
\bea
\frac{1}{i}\Delta'_c(p^2) & = &
 \frac{1}{i}\frac{1}{M^2-p^2-i\epsilon}+\frac{1}{i}\frac{1}{M^2-p^2-i\epsilon}i\Sigma(p^2)\frac{1}{i}\frac{1}{M^2-p^2-i\epsilon}+\ldots
 \nn \\ & = & \frac{1}{i}\frac{1}{M^2-p^2-i\epsilon-\Sigma(p^2)} \nn \\ & = &
 \frac{1}{i} \left[ \frac{Z_{\pi}}{M_{\pi}^2-p^2-i\epsilon}+R (p^2) \right] \;
 .
\label{tpf2}
\eea
In the last step we have developed the selfenergy function $\Sigma(p^2)$ in a
Taylor series around the physical mass $M_{\pi}^2$,
\bea
\Sigma(p^2) & = &
\Sigma(M_{\pi}^2)-(M_{\pi}^2-p^2)\bar{\Sigma}(M_{\pi}^2)+(M_{\pi}^2-p^2)^2
\stackrel{=}{\Sigma} (p^2) \; , \nn \\[0.3cm] 
{\bar{\Sigma}}(M_{\pi}^2) & = &
\frac{\partial}{\partial p^2}\Sigma(p^2)|_{p^2=M_{\pi}^2} \; , \nn \\ R(p^2) &
= & -\frac{ \stackrel{=}{\Sigma} (p^2)}{(1+\bar{\Sigma})
\,[\,1+\bar{\Sigma}-(M_{\pi}^2-p^2) \stackrel{=}{\Sigma} (p^2) \,] } \; ,
\label{tpf3}
\eea
where $Z_{\pi}=(1+\bar{\Sigma}(M_{\pi}^2))^{-1}$ is the residuum of the
propagator. The physical mass is defined as the position of the pole of the
propagator and we therefore have the implicit equation
\be
M^2-\Sigma(M_{\pi}^2) = M_{\pi}^2 \; ,
\label{tpf4}
\ee
which defines the physical mass $M_{\pi}$ order by order.

\begin{figure}[tb]
\unitlength1cm
\begin{picture}(1,1) \end{picture}
\epsfysize=3.5cm \epsffile{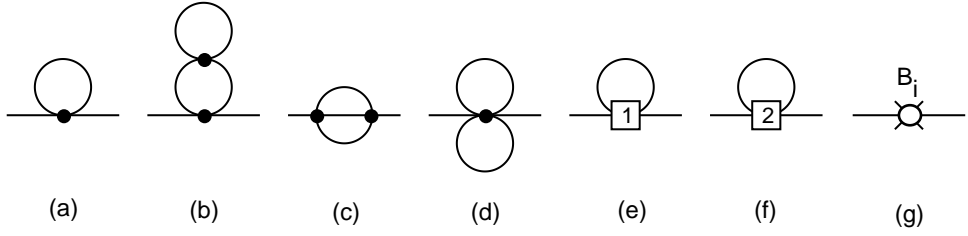}
\caption[Selfenergy contributions to the pion
propagator]{\label{mass}\small{Selfenergy-- graphs up to ${\cal
O}(p^6)$.
The contributions from diagram (a) start  at ${\cal O}(p^4)$,
whereas diagrams (b--g)
are two--loop corrections. The sunset diagram (c)
 contributes
to $R(p^2)$. The boxes in the graphs (e) and (f) denote vertices from the
lagrangian ${\cal L}_4$, where the numbers indicate the corresponding
low--energy constant. Diagram (g) contains the contribution from the lagrangian
${\cal L}_6$ only. }}
\end{figure}

Evaluating diagram (a) in Fig. \ref{mass} with the mass $M_0^2=M^2+2 M^4
l_3/F^4$ in the propagator and solving Eq. (\ref{tpf4}) one obtains the
physical pion mass up to ${\cal O}(p^4)$ given in Eq. (\ref{amp3}). The wave
function renormalization constant $Z_{\pi}$ and the finite part of the
propagator $R(s)$ become at this order
\be
Z_{\pi} = 1-\frac{1}{F^2} F_1(M^2) \;\; , \;\; R (s) = 0 \;\; , \;\;\;\; [\;
{\cal O}(p^4) \; ]
\label{tpf5}
\ee
where $F_1(M^2)$ is a one--loop function defined in Eq. (\ref{fm3}). At ${\cal
O}(p^6)$ the graphs (b--g) in Fig. \ref{mass} contribute to $\Sigma$. 
The diagrams (b--d) stem from the lagrangian ${\cal L}_2$ whereas (e)
and (f) are contributions from ${\cal L}_4$. The lagrangian ${\cal L}_6$
contributes via diagram (g). Only the sunset graph (c) is nontrivial. The overlapping
loop momenta in one of the propagators make the algebraic part
tedious. Nevertheless for $s=M_{\pi}^2$ the result of the sunset selfenergy
contribution may be obtained analytically.  We find for the pion mass up to
${\cal O}(p^6)$
\bea
M_{\pi}^2 & = & M^2 \left\{ 1+\frac{M^2}{F^2} \left(\frac{F_1(M^2)}{2 M^2}+2
                   l_3\right) \right.  \nn \\ & & \;\;\;\;\;\;\;\;\; +
                   \left. \frac{M^4}{F^4} \left[ F_2(M^2) \left( F_2(M^2)
                   p_M+\sum_{i=1}^3 l_i p_M^i \right) + B_M \right ] \right\}
                   + {\cal O}(M^8) \; .
\label{tpf6}
\eea
$B_M$ collects the contributions from the ${\cal L}_6$ lagrangian and
the polynomial $p_M^i$ read
\bea
p_M & = & \frac{17}{8}-\frac{79}{12}\omega+\frac{421}{32}\omega^2 + {\cal
O}(\omega^3) \; , \nn \\ p_M^1 & = &\frac{}{} 14-15\omega+\frac{31}{2}\omega^2
+ {\cal O}(\omega^3) \; , \nn \\[0.2cm] p_M^2 & = & \frac{}{}
8-10\omega+11\omega^2 + {\cal O}(\omega^3) \; , \nn \\[0.2cm] p_M^3 & = &
\frac{}{} 3-4\omega+4\omega^2 + {\cal O}(\omega^3) \; .
\label{tpf7}
\eea
$F_1,F_2$ are one--loop functions defined in Eq. (\ref{fm3}). For the wave
function renormalization constant $Z_{\pi}$ we get at ${\cal O}(p^6)$
\be
Z_{\pi} = 1-\frac{F_1(M^2)}{F^2}+\frac{M^4}{F^4} \left\{ F_2(M^2) \left(
F_2(M^2) p_Z+\sum_{i=1}^3 l_i p_Z^i \right) +B_Z \right\} +{\cal O}(M^6) \; ,
\label{tpf8}
\ee
where $B_Z$ collects the contributions from the ${\cal L}_6$ lagrangian and
where
\bea
p_Z & = & 1-\frac{29}{6}\omega+\frac{395}{32}\omega^2 + {\cal O}(\omega^3) \;
, \nn \\ p_Z^1 & = & \frac{}{}14-15\omega+\frac{31}{2}\omega^2 + {\cal
O}(\omega^3) \; , \nn \\[0.2cm] p_Z^2 & = & \frac{}{} 8-10\omega+11\omega^2 +
{\cal O}(\omega^3) \; , \nn \\[0.2cm] p_Z^3 & = & \frac{}{} 2 + {\cal
O}(\omega^3) \; .
\label{tpf9}
\eea
Furthermore $R(s)$ becomes
\bea
R(s) & = & \frac{F_2^2(M^2)}{F^4} p_R + \frac{1}{6 F^4} \frac{h_F(s)}{(s-M^2)^2} \; , \nn \\ p_R & =
& -\frac{3}{4}
M^2+(\frac{5}{4}M^2-\frac{1}{12}s)\omega-(\frac{53}{16}M^2+\frac{5}{32}s)\omega^2
+ {\cal O}(\omega^3) \; ,
\label{tpf10}
\eea
with
\bea
h_F(s) & = & \int_{4 M_{\pi}^2}^{\infty} d\sigma \beta(\sigma) \int_0^1 dx
\{\; 6 s M^4 +(1-12x+18x^2) s^2 M^2 \} {\cal K}(x,s) \; , \nn \\ {\cal K}(x,s) & =
& -\frac{1}{(16 \pi^2)^2}\left\{ \ln \frac{z_2(x,s)}{z_2(x,M^2)}+\frac{(s-M^2)
x (1-x)}{z_2(x,M^2)} \right\} \; , \nn \\[0.2cm] z_2(x,s) & = &
[\;M^2-s(1-x)\;]\;x + \sigma(1-x) \; ,
\label{tpf11}
\eea
and where $\beta(\sigma)=\sqrt{1-4 M^2/\sigma}$. The loop function
$h_F$ stems from the sunset diagram (c) in Fig. \ref{mass}.

\subsection{Contributions with one vertex from ${\cal L}_4$}

\begin{figure}[t]
\unitlength1cm
\begin{picture}(2,2) \end{picture}
\epsfysize=10cm \epsffile{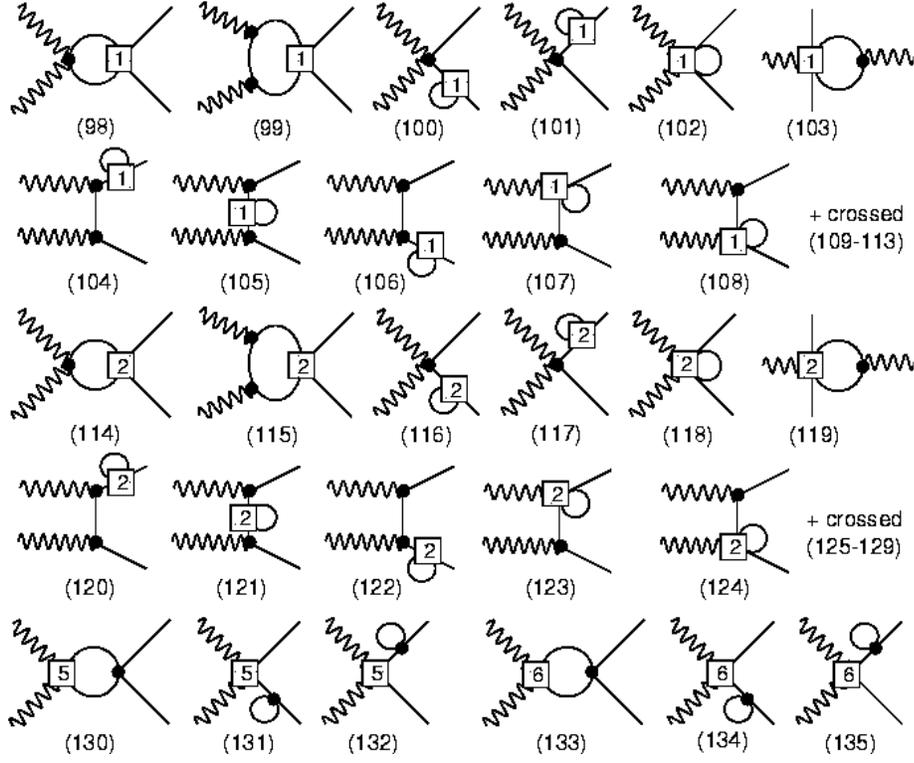}
\caption[One--loop diagrams with one vertex from ${\cal
L}_4$]{\label{1loopl4}\small{One--loop order contributions generated by the
lagrangians ${\cal L}_2+{\cal L}_4$ with one vertex from ${\cal L}_4$. The
boxes denote ${\cal L}_4$-- couplings, whereas the numbers indicate the
corresponding low--energy constant.}}
\end{figure}

We discuss some aspects of one--loop graphs generated by ${\cal L}_2+{\cal
L}_4$ with one vertex from the lagrangian ${\cal L}_4$ depicted in
Fig. \ref{1loopl4}. Only diagrams proportional to $(l_1,l_2,l_5,l_6)$ contribute to the
process $\ggpp$. The low energy constants $l_3$ and $l_4$ enter the two--loop
amplitude through mass-- and pion decay constant corrections. The diagrams
(98,99) and (114,115) in Fig. \ref{1loopl4} may be generated according to
Fig. \ref{pipi} where the ${\cal L}_4$--vertex is contained in the
$d$--dimensional elastic $\pi\pi$--scattering amplitude at one--loop
accuracy. These graphs remove the subdivergences indicated in Fig. \ref{pipi}
(enclosed by a dash--dotted line in graphs (22,36) ). \\ The diagrams (100,101)
and (116,117) may be included in the replacement $Z_{\pi}\otimes$ seagull and
the graphs (104--113), (120--129) are included in the replacement $M^2
\rightarrow M_{\pi}^2$ in the reducible Born diagrams. The remaining graphs
require straightforward one--loop calculation.

\subsection{Counterterms with one vertex from ${\cal L}_6$}

\begin{figure}[t]
\unitlength1cm
\begin{picture}(1,1) \end{picture}
\epsfysize=5cm \epsffile{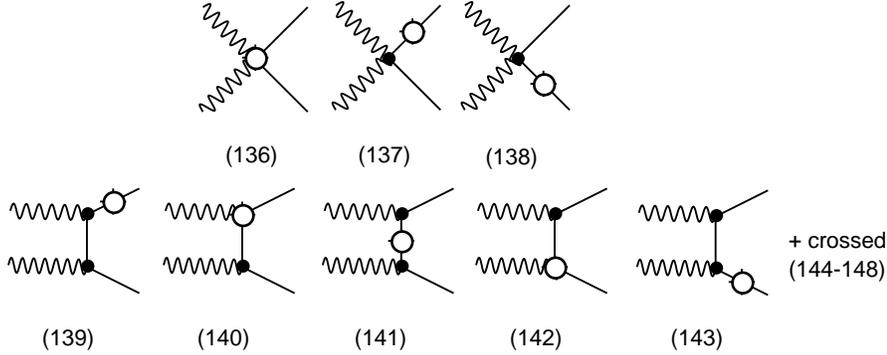}
\caption[Counterterms with one vertex from ${\cal
L}_6$]{\label{treel6}\small{Tree order counterterms generated by ${\cal
L}_2+{\cal L}_4+{\cal L}_6$ with one vertex from ${\cal L}_6$ denoted by a
white circle. Diagrams (137--139,141,143) denote mass corrections in the propagators.}}
\end{figure}

The diagrams depicted in Fig. \ref{treel6} and generated by the lagrangian
${\cal L}_6$ contribute a polynomial piece only. The divergent parts of the
couplings cancel the pole structure generated by the two--loop diagrams. We
have estimated the finite parts of these couplings by resonance exchange (see
below).  

This concludes the discussion of the diagrams which occur at
two--loop order in $\ggpp$. For a complete mathematical treatment of 
all genuine graphs occurring in $\ggpp (\ggppn)$ see \cite{BU96a,BU96b}.

\setcounter{equation}{0} \setcounter{subsection}{0}

\section{Checks \label{section8}}
The tensorial decomposition of the amplitude (\ref{kin4}) leads to formfactors
$c_i(s,\nu)$ (see \cite{BE94}), which are linearly related through
two Ward identities and through Bose symmetry,
\bea
L_1(c_i) & \doteq & 2 c_0+s c_4-(t-u) c_5 = 0 \; , \nn \\ L_2(c_i) & \doteq &
\hspace{1.0cm} s c_7-(t-u) c_9 = 0 \; , \nn \\ L_3(c_i) & \doteq &
\hspace{2.3cm} c_5 + c_7 = 0 \; .
\label{check1}
\eea
The amplitudes $A^{\sss{C}}$ and $B^{\sss{C}}$ may be obtained from $c_4$ and
$c_9$,
\be
A = -c_4 \; , \;\;\;\; B = c_9/2 s \; .
\label{check2}
\ee
We have evaluated the five formfactors $c_0,c_4,c_5,c_7$ and $c_9$ in $d$ dimensions and afterward the
following consistency checks were made:
\begin{description}

\item{(i) } We have checked that the relations (\ref{check1}) below threshold
are satisfied numerically within machine accuracy in low dimensions.

\begin{table}[t]
\caption[Pole coefficients of the two--loop
amplitudes]{\label{tab2}\small{Coefficients of the amplitude expansion in
Eq. (\ref{check3})}}\vspace{0.1cm}
\begin{center}
\begin{tabular}{|c||c|c|c|c|c|}  \hline
   & $i=0$ & $i=4$ & $i=5$ & $i=7$ & $i=9$ \\ \hline $a_0^i$ &
   $-\frac{2}{9}M^2s+\frac{1}{18}s^2-\frac{1}{18}\nu^2$ &
   $\frac{4}{9}M^2-\frac{1}{9}s$ & $-\frac{1}{9}\nu$ & $\frac{1}{9}\nu$ &
   $\frac{1}{9}s$ \\ $a_1^i$ &
   $\frac{337}{216}M^2s-\frac{127}{864}s^2-\frac{53}{864}\nu^2$ &
   $-\frac{337}{108}M^2+\frac{127}{432} s$ & $-\frac{53}{432}\nu$ &
   $\frac{53}{432}\nu$ & $\frac{53}{432}s$ \\ \hline $b_0^{i1}$ &
   $\frac{10}{3}M^2 s+\frac{1}{6}s^2-\frac{1}{6}\nu^2$ &
   $-\frac{20}{3}M^2-\frac{1}{3}s$ & $-\frac{1}{3}\nu$ & $\frac{1}{3}\nu$ &
   $\frac{1}{3}s$ \\ $b_1^{i1}$ & $-4M^2 s$ & $8 M^2$ & $0$ & $0$ & $0$ \\
   \hline $b_0^{i2}$ & $-3M^2 s+\frac{1}{4}s^2-\frac{1}{4}\nu^2$ &
   $6M^2-\frac{1}{2}s$ & $-\frac{1}{2}\nu$ & $\frac{1}{2}\nu$ & $\frac{1}{2}s$
   \\ $b_1^{i2}$ & $2 M^2 s$ & $-4 M^2$ & $0$ & $0$ & $0$ \\ \hline $b_0^{i5}$
   & $8M^2 s+2s^2$ & $-16M^2-4s$ & $0$ & $0$ & $0$ \\ $b_1^{i5}$ &$-8M^2 s$ &
   $16M^2$ & $0$ & $0$ & $0$ \\ \hline $b_0^{i6}$ & $-4M^2 s-s^2$ & $8M^2+2s$
   & $0$ & $0$ & $0$ \\ $b_1^{i6}$ & $4M^2 s$ & $-8M^2$ & $0$ & $0$ & $0$ \\
   \hline
\end{tabular}
\end{center}
\end{table}

\item{(ii)} By letting $\omega \rightarrow 0$ and omitting the tree contributions generated
by ${\cal L}_2+{\cal L}_4+{\cal L}_6$ in Fig. \ref{treel6}, the formfactors $c_i$ at order $p^6$ 
have the structure
\bea
c_i(s,\nu) & = & \frac{F_2(M^2)}{F^4} \left\{ F_2(M^2)
 \,[\,a_0^i+a_1^i\omega+a_2^i \omega^2+{\cal O}(\omega^3) \,] \right. \nn \\ &
 & \hspace{1cm}+\sum_{j=1,2,5,6} \left. l_j
 \,[\,b_0^{ij}+b_1^{ij}\omega+b_2^{ij}\omega^2+{\cal O}(\omega^3)
 \,]\,\right\} \nn \\ & = & \mu^{4\omega} \left\{
 \frac{P_i^{(2)}}{4\omega^2}+\frac{P_i^{(1)}}{2\omega}+R_i+{\cal O}(\omega)
 \right\} \; .
\label{check3}
\eea
where $a_k^i$ and $b_k^{ij}\; , k= 0,1$ are polynomials in the external
momenta and the pion mass (Table \ref{tab2}) and $F_2(M^2)$ is a loop function
defined in Eq. (\ref{fm3}).  The coefficients $a_2^i$ and $b_2^{ij}$ contain a
polynomial part as well as complicated analytic functions and contribute to
the finite part $R_i$ of the amplitude. \\ We have verified that the
residues $P_i^{(k)}$ are polynomials in the external momenta and the pion
mass. These polynomials obey Eq. (\ref{check1}) as well,
\be
L_m(P_i^{(k)}) = 0 \; , \;\;\;\; m=1,2,3 \; , \;\;\;\; k=1,2 \; .
\label{check4}
\ee
We have checked
numerically that
\be
L_m(R_i) = 0 \; , \;\;\;\; m=1,2,3 \; ,
\label{check5}
\ee
within machine accuracy below threshold. To go to the physical
region for pion--pair production we have written fixed--$t$ dispersion relations
for the box-- and vertex parts. Matching the dispersive analysis with the
non--dispersive expressions below threshold ensures that $L_m(R_i)=0$ is also satisfied above
threshold.

\item{(iii)} We have worked out the S--wave projection $h^0_{\sss{C},+}(s)$
(\ref{kin85}) for the helicity amplitude $H^{\sss{C}}_{++}$ and we have verified numerically that
these amplitudes admit the correct phase at $s>4 M_{\pi}^2$, given by the
(tree+one--loop) elastic $\pi\pi$ scattering S--wave phase shifts (in the
appropriate isospin decomposition).

\end{description}

\setcounter{equation}{0} \setcounter{subsection}{0}

\section{Amplitude and cross section to two loops \label{ampsec}}

\subsection{Analytic results of the amplitude}
We find for the two--loop amplitude $A^{\sss{C}} $
\be
A^{\sss{C}} = A^{\sss{C}}_2 + A^{\sss{C}}_4 + A^{\sss{C}}_6 + {\cal O}(p^4) \;
,
\label{2amp2}
\ee
or
\be
A^{\sss{C}}= \{\frac{1}{M_{\pi}^2-t}+\frac{1}{M_{\pi}^2-u}\}
      +\frac{2}{F_{\pi}^2}\{\bar{G}_{\pi}(s)+\frac{\bar{l}_{\Delta}}{48 \pi^2}
      \} + U^{\sss{C}}_{\sss{A}} + P^{\sss{C}}_{\sss{A}} + {\cal O}(p^4)\; .
\label{2amp3}
\ee
The unitary part $U^{\sss{C}}_{\sss{A}}$ contains $s$, $t$ and $u$-- channel
cuts, and $P^{\sss{C}}_{\sss{A}}$ is a linear polynomial in $s$. Explicitly we
find,
\bea
U^{\sss{C}}_{\sss{A}} & = & \frac{1}{s F_{\pi}^4} \bar{G}(s) \; [ \; (2
                    M_{\pi}^4-4 M_{\pi}^2 s + 3 s^2) \bar{J}(s) +
                    C^{\sss{C}}(s,\bar{l}_i) \; ] \;
                    +\frac{\bar{l}_{\Delta}}{48 \pi^2 F_{\pi}^4} s \bar{J}(s)
                    \nn \\ & & + \frac{(\bar{l}_1-\frac{4}{3})}{288 \pi^2 s
                    F_{\pi}^4}(s-4 M_{\pi}^2) \{\bar{H}(s)+4
                    \;[\;s\bar{G}(s)+2
                    M_{\pi}^2(\stackrel{=}{G}(s)-3\stackrel{=}{J}(s)) \;]\;
                    d_{00}^2 \} \nn \\ & & +\;
                    \frac{(\bar{l}_2-\frac{5}{6})}{96 \pi^2 s F_{\pi}^4} \;
                    (s-4 M_{\pi}^2) \{\bar{H}(s)+4 \;[\;s\bar{G}(s)+2
                    M_{\pi}^2(\stackrel{=}{G}(s)-3\stackrel{=}{J}(s)) \;]\;
                    d_{00}^2 \} \nn \vspace{.5cm} \\ & & +
                    \;\Delta^{\sss{C}}_{\sss{A}}(s,t,u) \; ,
\label{2amp4}
\eea
with
\bea
C^{\sss{C}}(s,\bar{l}_i) & = & \frac{1}{48 \pi^2} \{\frac{1}{3}(\bar{l}_1-\frac{4}{3})(16 s^2-56 M_{\pi}^2 s +64 M_{\pi}^4)
                                                                                      +(\bar{l}_2-\frac{5}{6})(8 s^2-24 M_{\pi}^2 s +32 M_{\pi}^4) \nn \\
                        & & -12 M_{\pi}^4 \bar{l}_3 +12 M_{\pi}^2 s \bar{l}_4-12 M_{\pi}^2 s +12 M_{\pi}^4 \} \; , \nn \\
d_{00}^2 & = & \frac{1}{2}(3\cos\theta^2-1) \; .
\label{2amp5}
\eea
The loop functions $\bar{J}$ etc. are displayed in appendix \ref{loopfunc} and
$\bar{G}_{\pi}(s)$ in Eq. (\ref{2amp3}) stands for $\bar{G}(s)$ evaluated with
the physical mass. The term proportional to $d_{00}^2$ in
$U^{\sss{C}}_{\sss{A}}$ contributes to $D$--waves only.
For
$\Delta^{\sss{C}}_{\sss{A}}$ see below. The polynomial part is
\bea
P^{\sss{C}}_{\sss{A}} & = & \frac{1}{(16\pi^2F_{\pi}^2)^2} \,[\,a^c_1
 M_{\pi}^2+a^c_2 s\,]\, \; , \nn \\ a^c_1 & = & a_1^{c,r}+\frac{1}{9}\{4
 l^2+l(-10\bar{l}_1+18\bar{l}_2-12\bar{l}_{\Delta}+\frac{337}{6})
 -\frac{5}{3}\bar{l}_1-5\bar{l}_2+12\bar{l}_4\bar{l}_{\Delta}+4 \} \; , \nn \\
 a^c_2 & = & a_2^{c,r}-\frac{1}{9}
 \{l^2+l(\frac{1}{2}\bar{l}_1+\frac{3}{2}\bar{l}_2+3\bar{l}_{\Delta}+\frac{127}{24})
 -\frac{5}{12}\bar{l}_1-\frac{5}{4}\bar{l}_2+3\bar{l}_{\Delta}+\frac{21}{2} \}
 \; , \nn \\ l & = & \ln \frac{M_{\pi}^2}{\mu^2} \; .
\label{2amp6}
\eea
The result for $B^{\sss{C}}$ reads
\be
B^{\sss{C}} = B^{\sss{C}}_2 + B^{\sss{C}}_6 + {\cal O}(p^2)\; ,
\label{2amp7}
\ee
or
\be
B^{\sss{C}} = \frac{1}{2s}\{\frac{1}{M_{\pi}^2-t}+\frac{1}{M_{\pi}^2-u}\} +
U^{\sss{C}}_{\sss{B}} + P^{\sss{C}}_{\sss{B}} + {\cal O}(p^2) \; ,
\label{2amp8}
\ee
with the unitary part
\bea
U^{\sss{C}}_{\sss{B}} & = & \frac{1}{192 \pi^2 s F_{\pi}^4} \{
                 \frac{1}{3}(\bar{l}_1-\frac{4}{3})+(\bar{l}_2-\frac{5}{6}) \}
                 \bar{H}(s) + \Delta^{\sss{C}}_{\sss{B}}(s,t,u) \; .
\label{2amp9}
\eea
For the polynomial part we find
\bea
P^{\sss{C}}_{\sss{B}} & = & \frac{b^c}{(16\pi^2 F_{\pi}^2)^2} \; , \nn \\ b^c
& = & b^{c,r}-\frac{1}{18} \{ l^2+l
(\frac{1}{2}\bar{l}_1+\frac{3}{2}\bar{l}_2-\frac{53}{24})-\frac{1}{12}\bar{l}_1
-\frac{1}{4}\bar{l}_2+\frac{7}{2} \} \; .
\label{2amp10}
\eea
The integrals $\Delta^{\sss{C}}_{\sss{A,B}}(s,t,u)$ contain contributions from
the two--loop box--, vertex-- and acnode graphs and also from the reducible
diagrams manifest in the function $R(s)$ given in Eq. (\ref{tpf10}). We
discuss the size of these corrections in section \ref{numeric}.

\subsection{Low--energy constants \label{lowcon}}
The Born contribution to the charged pion--pair production contains $M_{\pi}$
as the only parameter. At next--to--leading order three parameters occur, namely,
 $M_{\pi}, F_{\pi}, \bar{l}_6- \bar{l}_5$. Going to the two--loop correction the
number of parameters increases. Once the program described in section
\ref{eval2loop} is carried through, the ultraviolet finite and scale
independent amplitudes $A^{\sss{C}}$ and $B^{\sss{C}}$ contain the set
$M_{\pi}, F; l_i^r,l_{i1}^r \; (i=1,2,3,5,6); a_1^{r,c},a_2^{r,c}$ and
$b^{r,c}$ of low--energy constants. The couplings $l_{i1}^r$ stem from the
${\cal O}(\omega)$--part in the expansion of the ${\cal L}_4$ low--energy parameters, 
whereas $a_1^{r,c},a_2^{r,c}$ and $b^{r,c}$ are couplings from the lagrangian ${\cal L}_6$.
 $F$ is related to the pion decay constant
$F_{\pi}$ \cite{GL84}
\be
F_{\pi} = F \left[ \,
1+\frac{M^2}{F^2}\left(l_4^r-\frac{1}{16\pi^2}\ln\frac{M^2}{\mu^2}\right)
+{\cal O}(M^4) \, \right] \; .
\label{lconst1}
\ee
We may therefore replace $F$ by $F_{\pi}$ at the expense of introducing
$l_4^r$. The expressions for the loop--amplitudes simplify if one uses the
scale--independent parameters $\bar{l}_i$ instead of $l_i^r$ ,
\be
l_i^r = \frac{\gamma_i}{32\pi^2}(\bar{l}_i+\ln\frac{M^2}{\mu^2})\; .
\label{lconst2}
\ee
The values of the $\bar{l}_i$ are depicted in Table \ref{tab1} together with
the experimental sources and the values for the $\gamma_i$. The couplings
$l_{i1}^r$ contribute a polynomial piece only and may therefore be absorbed
into the low--energy constants at order $p^6$ \cite{BI96}.
\begin{table}[t]
\begin{center}
\caption[Low--energy constants]{\label{tab1}\small{Low--energy constants and
sources. In the fourth column we have plotted the $\gamma_i$ evaluated in
\cite{GL84,BC94}}}\vspace{0.1cm}
\begin{tabular}{|c||r|l|c|}  \hline
  i & $\bar{l}_i$ & Source & $\gamma_i$ \\ \hline 1 & $-1.7 \pm 1.0$ &
  $K_{e4}, \pi\pi \rightarrow \pi\pi$ & $1/3$ \\ 2 & $ 6.1 \pm 0.5$ & $K_{e4},
  \pi\pi \rightarrow \pi\pi$ & $2/3$ \\ 3 & $ 2.9 \pm 2.4$ & $SU(3)$ mass
  formulae & $-1/2$ \\ 4 & $ 4.3 \pm 0.9$ & $F_K/F_{\pi}$ & $2$ \\ 5 & $ 13.8
  \pm 1.3$ & $\pi \rightarrow e \nu \gamma$ & $-1/6$ \\ 6 & $ 16.5 \pm 1.1$ &
  $<r^2>_V^{\pi}$ & $-1/3$ \\ \hline
\end{tabular}
\end{center}
\end{table}
We are therefore left with $a_1^{r,c}, a_2^{r,c}$ and $b^{r,c}$ as the only
new unknowns. The coupling constants occurring at ${\cal O}(p^4)$ in the
low--energy expansion are dominated by the low--lying vector, axial--vector,
scalar and pseudoscalar resonances in such a way, that they
practically saturate the corresponding couplings of the ${\cal O}(p^4)$
effective lagrangian \cite{GL84, EG89}. 

Here we assume that this is also true at order $p^6$. We estimate the
renormalized couplings $a_1^{r,c}, a_2^{r,c}$ and $b^{r,c}$ by replacing them at
a scale $\mu=500 MeV \ldots 1 GeV$ by the contribution from resonance
exchange.  Let
\be
I^{c,r}(\mu) = \sum_{R=\rho,a_1,b_1} \, I^R +\hat{I}^r(\mu) \; , \;\;\;\;
I=a_1,a_2,b \; ,
\label{lowc3}
\ee
where the sum denotes contributions from vector-- $(J^{PC}=1^{--})$ and
axial--vector $(J^{PC}=1^{++}, 1^{+-})$ exchange \cite{DH93,PE81,KA86,KO93,BA93}
with meson mass $M_R \le 1.2$
GeV. The non--resonance contribution of these low--energy constants has to be
chosen in such a way that the final result of the amplitudes $A^{\sss{C}}$ and
$B^{\sss{C}}$ is scale independent. Our estimate for $I^{c,r}(M_{\rho})$ consists
in setting $\hat{I}^r(M_{\rho})=0$.

\begin{table}[t]
\begin{center}
\caption[Resonances]{\label{resoest}\small{Resonance contributions to the
constants $a_1^{r,c},a_2^{r,c}$ and $b^{r,c}$. The calculation is done in \cite{BU96a},
the uncertainties are more generous than quoted in that reference.
}}\vspace{0.2cm}
\begin{tabular}{|c||c|c|c||c|}  \hline
  $I^r$ &\multicolumn{3}{|c|}{$I^R$} & $\sum_R I^R$ \\ \cline{2-4} & $\rho$ &
$a_1$ & $b_1$ & \\ \hline $a_1^{r,c}$ & -3.28 &0 &0 & $-3.3 \pm 1.65$ \\
$a_2^{r,c}$ & 1.23 & -0.35 & -0.13 & $0.75 \pm 0.65$ \\ $b^{r,c}$ & 0.20 &0.18
& 0.06 & $0.45 \pm 0.15$ \\ \hline
\end{tabular}
\end{center}
\end{table}

The quantities $I^R$ are evaluated in \cite{BU96a}. The results
of this calculation are displayed in Table \ref{resoest} where the individual
resonance contributions $I^R$ are listed. Column 5 contains the sums of the
several contributions provided with a systematic error. We associate a $50\%$
uncertainty to the contributions generated by (axial--) vector exchange.

It is useful to consider furthermore the helicity amplitudes
$H^{\sss{C}}_{++}$ and $H^{\sss{C}}_{+-}$ and the corresponding low--energy
constants $h^{r,c}_{\pm}$ and $h^{r,c}_s$,
\bea
H_{++}^{{\sss{C}},2 loops} & = & \frac{h_+^{r,c} M^2 + h_s^{r,c} s}{(16\pi^2
F^2)^2} + \ldots \nn \; , \\ H_{+-}^{{\sss{C}}, 2 loops} & = & \frac{8(M^4-t
u)}{s (16\pi^2 F^2)^2} h_-^{r,c} + \ldots \; , \nn \\ h_+^{r,c} & = &
a_1^{r,c}+8 b^{r,c} \; , \;\;\; h_s^{r,c} = a_2^{r,c}-2 b^{r,c} \; , \;\;\;
h_-^{r,c}=b^{r,c} \; .
\label{lowc14}
\eea
Adding the errors quoted in Table \ref{resoest}  in quadrature, we find
\bea
h_+^{r,c}(M_{\rho}) & = & \;\;\;0.3 \pm 2.0 \; , \nn \\ h_s^{r,c}(M_{\rho}) &
= & - 0.15 \pm 0.7\; , \nn \\ h_-^{r,c}(M_{\rho}) & = & \;\;\;0.45 \pm 0.15\; .
\label{lowc15}
\eea
This completes the determination of the parameters which occur at two--loop
order in $\ggpp$.

\subsection{Numeric results of the amplitude \label{numeric}}
\begin{figure}
\unitlength1cm
\begin{picture}(0.5,1) \end{picture}
\epsfysize=8.8cm \epsffile{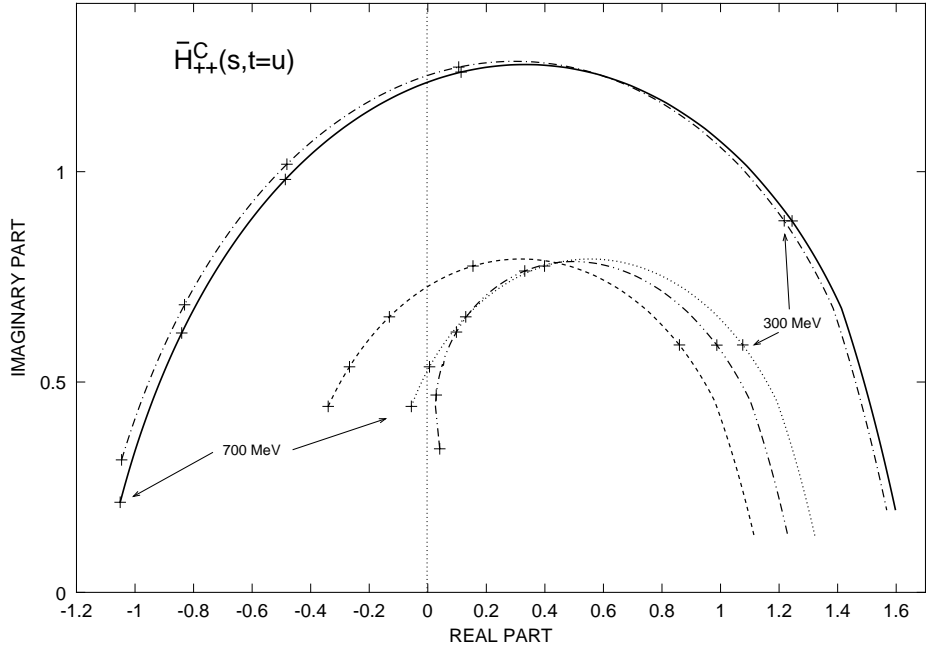} \\[0.5cm]
\begin{picture}(0.5,1) \end{picture}
\epsfysize=8.8cm \epsffile{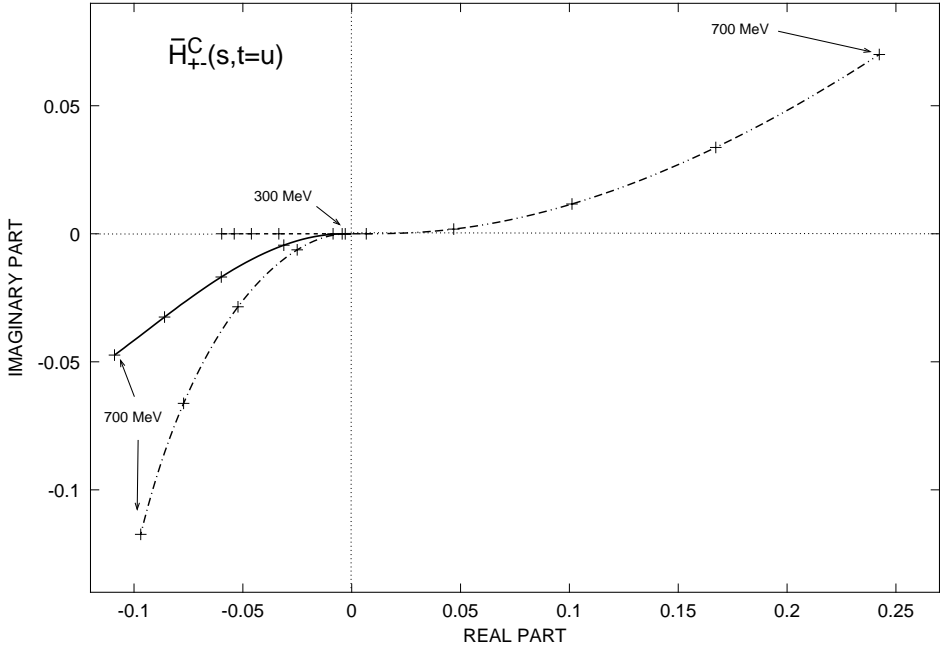}
\caption[Helicity amplitudes $H^{\sss{C}}_{+\pm}$]{\label{helampl}\small{Real
and imaginary part of the helicity amplitude $10^{-3} M_{\pi}^2
\bar{H}^{\sss{C}}_{+\pm}$ at $t=u$. The solid line incorporates
all contributions from one-- and two--loop corrections. The dash--dotted line is
the same amplitude with the omission of  $\Delta^{\sss{C}}_{A,B}$. The dotted line
corresponds to the one--loop result
(apart from mass renormalization terms, $\bar{H}^{\sss{C}}_{+-}$ does not receive any
contribution
from one--loop diagrams -- the dotted line is therefore absent in the lower figure). 
The dashed line is the sum of the one--loop and
acnode, the dashed--double dotted the sum of the one--loop and box contribution
at the scale $\mu=770$ MeV in the $\overline{MS}$  chiral scheme. Finally the
crosses refer to the center--of--mass energy of the $\pi^+\pi^-$ system in $100$
MeV steps. }}
\end{figure}

The characteristics of the two--loop corrections may be seen in
Fig. \ref{helampl}, where we have plotted the real and imaginary part of the
helicity amplitudes $10^{-3} M_{\pi}^2 \bar{H}^{\sss{C}}_{+\pm}$ (\ref{kin86}) at $t=u$. 
All curves start at
$E=280$ MeV and the crosses refer to the center--of--mass energy of the
$\pi^+\pi^-$ system in $100$ MeV steps \footnote{We use $F_{\pi}=92.4$ MeV,
$M_{\pi}=139.6$ MeV and $\bar{l}_1=-1.7\pm 1.0$, $\bar{l}_2= 6.1 \pm 0.5$,
$\bar{l}_3= 2.9 \pm 2.4$, $\bar{l}_4=4.3 \pm 0.9 $, $\bar{l}_{\Delta} = 2.7 \pm 0.4$
throughout.}. The solid line incorporates all
contributions stemming from the one-- and two--loop calculation only. The
dash--dotted line is the same amplitude without $\Delta^{\sss{C}}_{A,B}$. The dotted line
corresponds to the one--loop result
(apart from mass renormalization terms, $\bar{H}^{\sss{C}}_{+-}$ does not receive any
contribution
from one--loop diagrams). To get some feeling about the size of the genuine
two--loop diagrams, we have drawn with a dashed line the contributions from the
sum of one--loop and acnode graphs, with a dashed--double dotted line the sum of
the one--loop and box diagrams at the scale $\mu=770$ MeV in the $\overline{MS}$
chiral scheme. 

\subsection{Cross section $\ggpp$ \label{crosssec}}

\begin{figure}[tb]
\unitlength1cm
\begin{picture}(0.5,1) \end{picture}
\epsfysize=9cm \epsffile{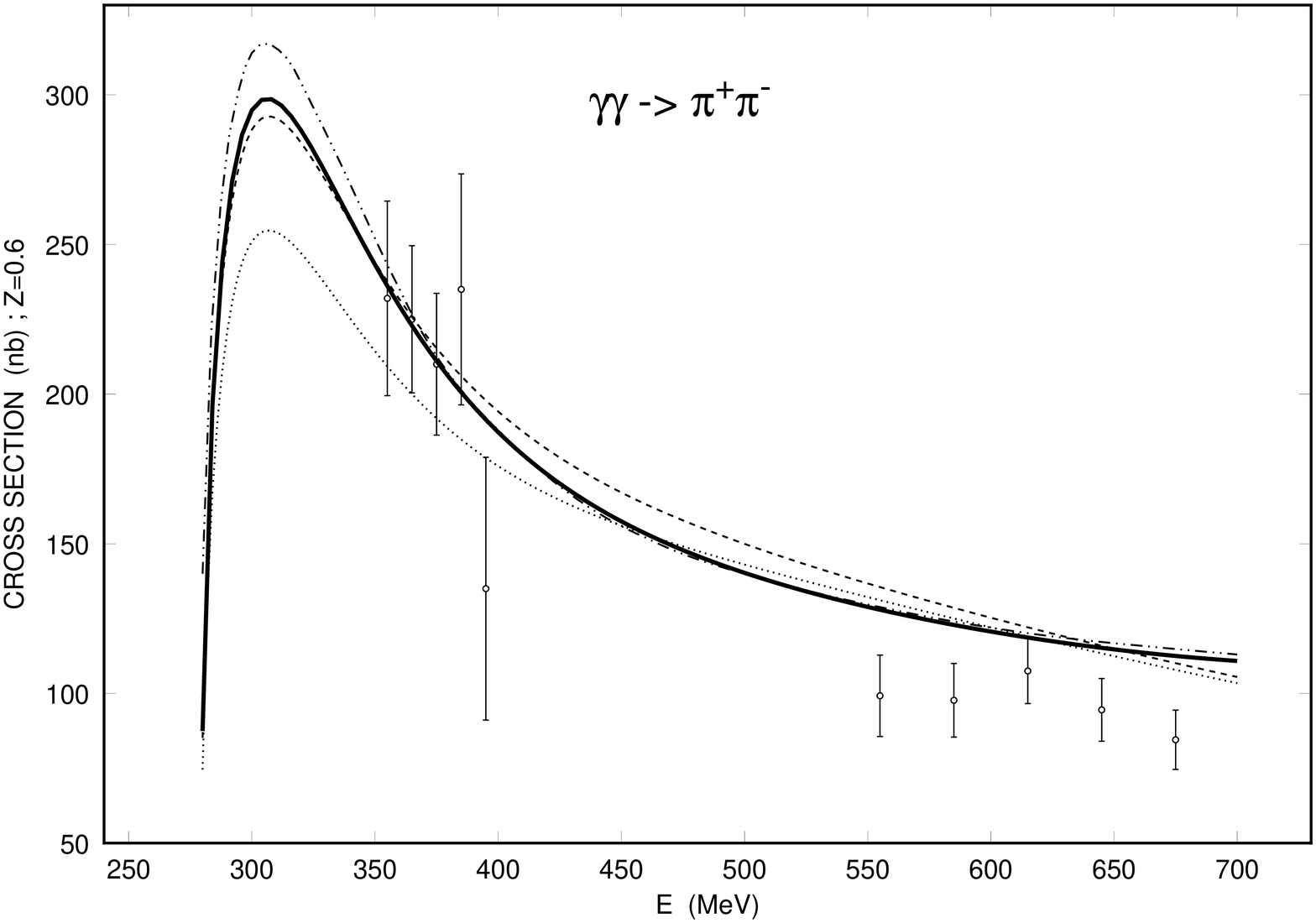}
\caption[Two--loop cross section $\ggpp$]{\label{ggpppm1}\small{The $\ggpp$
cross section $\sigma(s\, ; \, |\cos\theta\,|\le Z=0.6)$ as a function of the
center--of--mass energy $E$, together with the data from the Mark II
collaboration \cite{mark}. We have added in quadrature the tabulated statistical 
and systematical errors. In addition, there is an overall normalization uncertainty 
of $7 \%$ in the data \cite{mark}. 
The solid line is the full two--loop result, the dashed line
corresponds to the one--loop approximation \cite{BC88} and the
dotted line is the Born contribution. The dashed--double dotted line is the
result of a dispersive calculation performed by Donoghue and Holstein (Fig. 7 in
Ref. \cite{DH93}).}}
\end{figure}

The cross section for $\ggpp$ in the framework of $SU(3)\times SU(3)$ CHPT was
worked out up to ${\cal O}(p^4)$ in \cite{BC88}. The amplitude at this order
contains only the three low--energy parameters $F_{\pi},M_{\pi}$ and
$\bar{l}_{\Delta}$. We note that the increase in the cross section around the
peak at $\sqrt{s}=306 MeV$ due to the one--loop correction is $\sim 15 \%$
where the renormalization scale independent counterterm contribution $\sim
\bar{l}_{\Delta}$ amounts $\sim 50 \%$ of this correction. Furthermore it is
instructive to compare the cross section with the one from $\ggppn$ scattering
\cite{BE94}. In the neutral case the size of the cross section is two orders
of magnitude smaller than in the charged channel, since the neutral amplitude
starts out with one--loop diagrams, whereas in the charged case the Born
amplitude contributes dominantly to the cross section: At low energies
the photon couples to the charge of the pions. As a result of this, the $\ggppn$ cross
section is small, while that for $\ggpp$ is large.

\begin{figure}[tb]
\unitlength1cm
\begin{picture}(0.5,1) \end{picture}
\epsfysize=9cm \epsffile{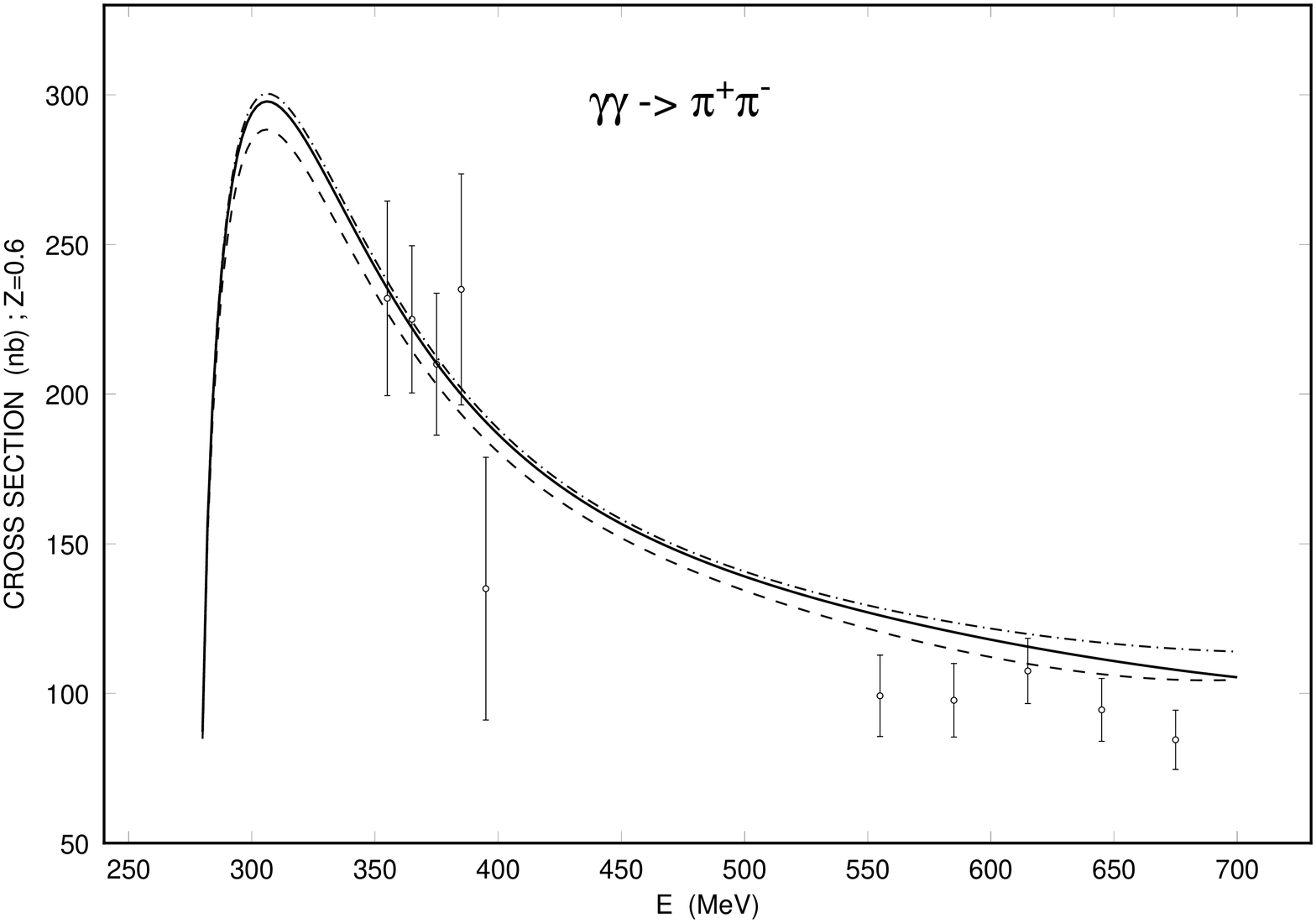}
\caption[Dependence of the two--loop cross section $\ggpp$ on the
$\bar{l}_i$]{\label{ggpppm2}\small{The $\ggpp$ cross section $\sigma(s\, ; \,
|\cos\theta\,|\le Z=0.6)$ as a function of the center--of--mass energy $E$. The
solid line is the two--loop result evaluated at $\Delta^{\sss{C}}_{A,B}=0$ and
without resonance exchange. The dashed line corresponds to $\bar{l}_i=0$,
whereas the dash--dotted line is obtained by setting
$\bar{l}_1=\bar{l}_3=0$. }}
\end{figure}

The plot in
Fig. \ref{ggpppm1} shows the total cross section $\sigma(s\, ; \, |\cos\theta\,
| \le Z=0.6)$ as a function of the center--of--mass energy $E=\sqrt{s}$. Here we
choose $Z=0.6$ to compare with the available data from the Mark II
collaboration \cite{mark}. The Born contribution is displayed with a dotted
line whereas the born+one--loop cross section is shown with a dashed line. Also
at this order the chiral expansion is in remarkable agreement with the
available data.  The solid line incorporates all contribution up to ${\cal
O}(p^6)$. It is seen that the two--loop effect changes the one--loop result only
very little. Switching off the integrals $\Delta^{\sss{C}}_{A,B}$ does not
affect the cross section visibly. This fact is also clear by consulting
Fig. \ref{helampl}, from where it is seen that our explicit, compact  representation of
the two--loop amplitudes is rather accurate. We note that
the two--loop correction, mainly dominated by the effect of final state $\pi\pi$ 
interaction, enhances the cross section very close to threshold and
then suppress it above $E=350 $ MeV, see Fig. \ref{ggpppm1}. Finally, the
dashed--double dotted line displays the result of a dispersive analysis (Fig. 7
in Ref. \cite{DH93}). In that calculation, use was made of a doubly subtracted
dispersion relation with the subtraction constants determined in terms of
chiral counterterms, incorporating also heavy meson exchange.

The two--loop result thus agrees very well with the data and shows furthermore
a very good agreement for $\sqrt{s} \ge 350$ MeV with the dispersive analysis
of Donoghue and Holstein.

In Fig. \ref{ggpppm2} the effect of the low--energy constants is shown. The
solid line corresponds to the two--loop result evaluated at
$\Delta^{\sss{C}}_{A,B}=0$ and without resonance exchange. (It turns out that
the contributions from the resonances to the cross section are negligible
below $450$ MeV.)
The dashed line corresponds to $\bar{l}_i=0$, and the dash--dotted line is
obtained by setting $\bar{l}_1=\bar{l}_3=0$. Therefore the increase in the
cross section is due to $\bar{l}_2, \bar{l}_4$ and $\bar{l}_{\Delta}$, as was
also shown in the neutral case \cite{BE94}.

\setcounter{equation}{0} \setcounter{subsection}{0}

\section{Compton scattering and pion polarizability \label{compsec}}

Compton scattering on pions $\gpgp$ is related to pion--pair production by crossing 
$(p_1,q_1)^{\mu} \rightarrow -(p_1,q_1)^{\mu},  s \leftrightarrow t $. The 
corresponding scattering amplitudes may again be expressed in terms of the
amplitudes $A^{\sss{C}}$ and $B^{\sss{C}}$ as functions of the Mandelstam 
variables in the Compton channel,
\be
\bar{s} = (q_1+p_1)^2 \; , \; \bar{t}=(q_2-q_1)^2 \; , \; \bar{u}=(p_2-q_1)^2 \; .
\label{kincom1}
\ee
In the present section we discuss shortly the Compton cross section.
Afterwards we work out the chiral expansion of the charged pion polarizabilities 
at next--to--leading order.

\subsection{Compton cross section $\gamma\pi^+ \rightarrow \gamma\pi^+$}

\begin{figure}[tb]
\unitlength1cm
\begin{picture}(0.5,1) \end{picture}
\epsfysize=9cm \epsffile{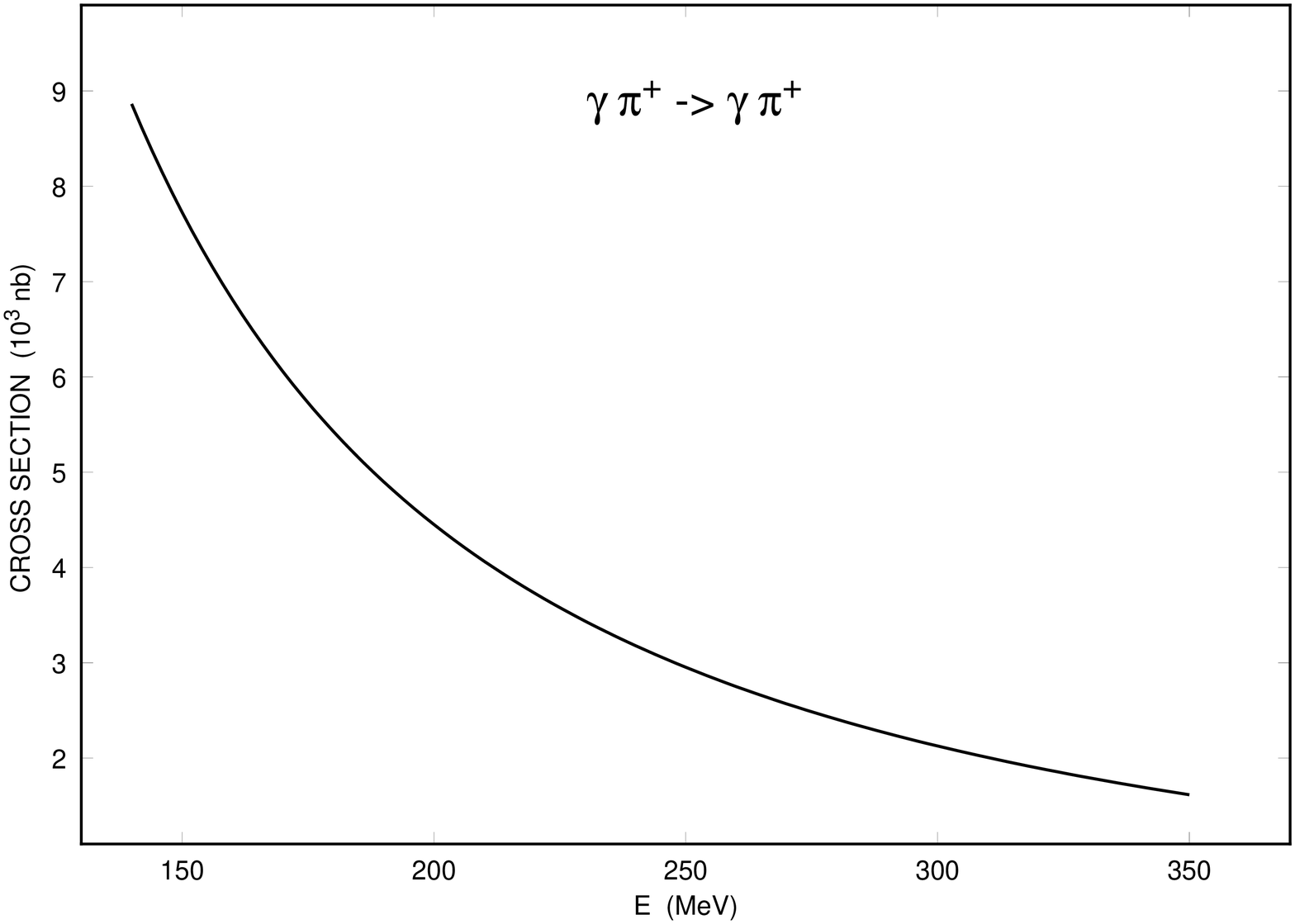}
\caption[Cross section
$\gamma\pi^+\rightarrow\gamma\pi^+$]{\label{gpgp}\small{The Compton
 cross section  as a
function of the center--of--mass energy $E=\sqrt{\bar{s}}$ in Born
approximation.  One-- and two--loop corrections are tiny in this case and would
not be visible in the figure. }}
\end{figure}

The total cross section at tree level is given by \cite{KO84}
\be
\sigma^{\gamma\pi^+ \rightarrow \gamma\pi^+} (\bar{s}) = 4 \alpha^2 \pi
\frac{M_{\pi}^2+\bar{s}}{\bar{s}(M_{\pi}^2-\bar{s})^3} \left[\,
M_{\pi}^4-\bar{s}^2+2 M_{\pi}^2\, \bar{s}\, \ln\frac{\bar{s}}{M_{\pi}^2} \,\right]
\; .
\label{comp1}
\ee
This quantity is plotted in Fig. \ref{gpgp} as a function of the center--of--mass
energy $E_{\gamma\pi}$. At threshold one obtains the total Thomson cross
section
\be
\sigma^{\gamma\pi^+ \rightarrow \gamma\pi^+} (\bar{s}=M_{\pi}^2) = \frac{ 8
\pi}{3} r_e^2 \; ,
\label{comp2}
\ee
where $r_e=\alpha/M_{\pi}$ is the classical electromagnetic radius of the pion. Comparing the
Compton cross section with the pion production case we note that the former is
one order of magnitude bigger. The increase in the cross section due to
one--loop corrections is smaller than one percent and the two--loop result is
again smaller by a factor of 3 . As a result of this, corrections in the
charged Compton channel are negligible.

In the neutral
channel,  the leading term is generated by one--loop graphs.
 The
cross section at $E=350$ MeV is three orders of magnitude smaller than in the
charged case and tends to zero at threshold.  Two--loop corrections are
substantial,
 and the result for the Compton amplitude is therefore not very reliable
at this order \cite{BE94}.

\subsection{Chiral expansion of the pion polarizabilities}

Among other fundamental parameters (mass, charge, magnetic moment, ...) the
electromagnetic polarizabilities characterize a composite system like a hadron
\cite{polar}.  They parametrize the first correction in the Taylor 
series expansion of the Compton amplitude in photon energies at threshold, 
\be
 T = -2 \left[ \vec{\epsilon}_1 \cdot \vec{\epsilon}_2\:\!^*(e^2-4 \pi
M_{\pi}\bar{\alpha}_{\pi} \omega_1\omega_2)-4 \pi M_{\pi}
\bar{\beta}_{\pi}(\vec{q}_1\times\vec{\epsilon}_1)\cdot(\vec{q}_2\times\vec{\epsilon}_2\:\!^*)
+ \dots \right ] \; ,
\label{pol3}
\ee
with $q_i^{\mu}=(\omega_i,\vec{q}_i)$. In terms of the helicity amplitudes $\bar{H}^{\sss{C}}_{++} (J=0)$ 
and $\bar{H}^{\sss{C}}_{+-}
(J=2)$ one has 
\be
\bar{\alpha}_{\pi^{\pm}} \pm \bar{\beta}_{\pi^{\pm}} =
\frac{\alpha}{M_{\pi}}\lim_{s \rightarrow 0} \lim_{t \rightarrow
M_{\pi}^2}\bar{H}^{\sss{C}}_{+\mp} \; ,
\label{pol5}
\ee
indicating that the combination
$\bar{\alpha}_{\pi^{\pm}}-\bar{\beta}_{\pi^{\pm}}$ is pure S--wave while
$\bar{\alpha}_{\pi^{\pm}}+\bar{\beta}_{\pi^{\pm}} $ is pure D--wave in the
$\gamma\pi^{\pm} \rightarrow \gamma\pi^{\pm}$ channel.  
Below we also use the
notation
\bea
\apmb & = & \bar{\alpha}_{\pi^{\pm}} \pm \bar{\beta}_{\pi^{\pm}} \; , \nn \\
\apmbn & = & \bar{\alpha}_{\pi^0} \pm \bar{\beta}_{\pi^0} \; .
\eea

From the Compton scattering amplitude at threshold Eq. (\ref{pol3}) expressed 
through the functions 
$A^{\sss{C}}$ and $B^{\sss{C}}$ (see Eqs. (\ref{2amp3}, \ref{2amp8})) it is
straightforward to determine the chiral expansion of the pion 
polarizabilities in terms of the quark masses to two loops,
\bea
(\alpha \pm \beta)^{\sss{N,C}} & = & \frac{\alpha}{16 \pi^2 M_{\pi} F_{\pi}^2}
\left\{ \; A_{\pm}^{\sss{N,C}} + \frac{M_{\pi}^2 B_{\pm}^{\sss{N,C}}}{16 \pi^2
F_{\pi}^2} + {\cal O}(M^4_{\pi}) \; \right\} \; .
\label{polchi1}
\eea
The leading result is given in \cite{HO90}
\be
A_{\pm}^{\sss{N}} = \left( \begin{array}{c} 0 \\ -\frac{1}{3} \end{array}
\right) \;\;\; , \;\;\; A_{\pm}^{\sss{C}} = \left( \begin{array}{c} 0 \\
\frac{2}{3} \bar{l}_{\Delta} \end{array} \right) \; .
\label{polchi2}
\ee
The next--to--leading order terms $B_{\pm}^{\sss{N}}$ have been determined in
\cite{BE94}. We find the following expressions for the coefficients
$B_{\pm}^{\sss{C}}$
\bea
B_+^{\sss{C}} & = & 8 h^{r,c}_-(\mu) -\frac{4}{9}
                                 \left\{l(l+\frac{1}{2}\bar{l}_1+\frac{3}{2}\bar{l}_2)-\frac{53}{24}l
                                 + \frac{1}{2}\bar{l}_1
                                 +\frac{3}{2}\bar{l}_2+\frac{91}{72} +
                                 \Delta_+^{\sss{C}} \right\} \;\; , \nn \\
                                 B_-^{\sss{C}} & = & h^{r,c}_+(\mu) -\frac{4}{3}
                                 \left\{l(\bar{l}_1-\bar{l}_2+\bar{l}_{\Delta}-\frac{65}{12})
                                 - \frac{1}{3}\bar{l}_1
                                 -\frac{1}{3}\bar{l}_2+\frac{1}{4}\bar{l}_3-\bar{l}_{\Delta}\bar{l}_4
                                 +\frac{187}{108} + \Delta_-^{\sss{C}}
                                 \right\} \; , \nn \\ l & = & \ln
                                 \frac{M_{\pi}^2}{\mu^2} \; ,
\label{polchi3}
\eea
with
\bea
\Delta_+^{\sss{C}} & = & -18 (16 \pi^2 F_{\pi}^2)^2 \;\;\;\;\; \lim_{s
\rightarrow 0} \lim_{t \rightarrow M_{\pi}^2} \dbc \hspace{2cm} = -8.69\; ,
\nn \\ \Delta_-^{\sss{C}} & = & -\frac{3}{4 M_{\pi}^2} (16 \pi^2 F_{\pi}^2)^2
\lim_{s \rightarrow 0} \lim_{t \rightarrow M_{\pi}^2}(\dac+8 M_{\pi}^2 \dbc) =
-8.73 \; .
\label{polchi4}
\eea
Our final result of the charged pion polarizabilities including corrections up
to ${\cal O}(p^6)$ reads $(\mu=M_{\rho})$
\bea
\apb & = & \;\;\; 0.3 \pm 0.1 \hspace{1cm} (0.0) \nn \; , \\ 
\amb & = &  \;\;\; 4.4 \pm 1.0 \hspace{1cm} (5.4\pm 0.8)\nn \; , \\ 
\bar{\alpha}_{\pi^{\pm}} & = & \;\;\; 2.4 \pm 0.5 \hspace{1cm} (2.7 \pm 0.4)\nn \; , \\ 
\bar{\beta}_{\pi^{\pm}} & = & -2.1 \pm 0.5 \hspace{1cm} (-2.7 \pm 0.4)\; .
\label{polchi7}
\eea
The numbers in brackets denote the leading order result. 
The estimate of the errors only stems from the uncertainties in the couplings
and do not contain effects from higher orders in the quark mass expansion or
any correlations. 
(The errors are more generous than quoted in Ref. \cite{BU96a}.)
We use furthermore $\bar{l}_{\Delta} = 2.7 \pm 0.4$ \cite{GL84}
extracted from the tree result $({\cal O}(p^4))$ of the radiative pion beta decay. 
Note however, that corrections in this process may change the error accordingly.

\subsection{Data on pion polarizabilities \label{polardata}}

\begin{figure}[t]
\unitlength1cm
\begin{picture}(1,1) \end{picture}
\epsfysize=5cm \epsffile{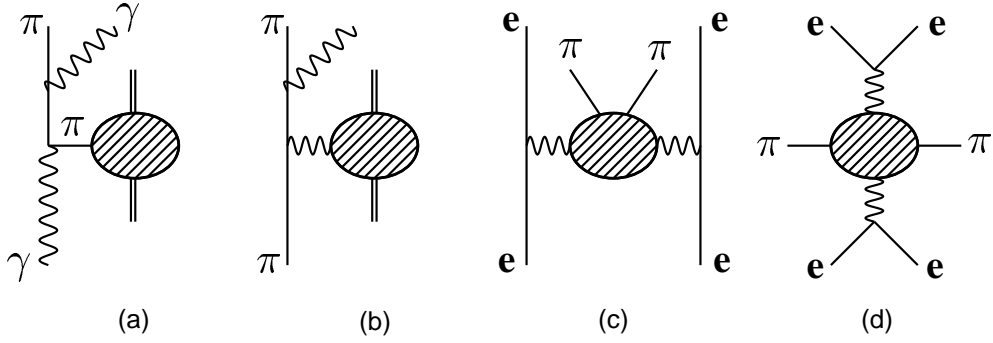}
\caption[Compton scattering and photon
annihilation]{\label{scatter}\small{Compton scattering $\gamma\pi \rightarrow
\gamma\pi$ from the charged pion has been performed both via pion--photon
production (a) and via radiative pion scattering (b) from a heavy
target. Figures (c) and (d) show photon annihilation
$\gamma\gamma\rightarrow\pi\pi$ processes. The double lines denote nucleons.}}
\end{figure}

Performing Compton scattering on pions is not easy. Nevertheless, it is
possible to probe the pion polarizability according to Fig. \ref{scatter} by
measuring the Compton scattering amplitude: (a) Pion photoproduction in
photon--nucleus scattering $\gamma p \rightarrow \gamma \pi^+ n$ \cite{AI86},
(b) Radiative pion nucleus scattering (Primakoff effect) $\pi^-- Z \rightarrow
\pi^-- \gamma Z$ \cite{AN84} or (c,d) Pion--pair production in $e^+e^-$--
collisions. Analyzing the data with the constraint $\apb = 0$ gives

\be
\amb = \left\{ \begin{array}{r@{\: \pm \:} l} 40 & 24 \;\;\;\; \mbox{(Lebedev
                       \cite{AI86} )} \; ,\\ 13.6 & 2.8 \;\;\;
                       \mbox{(Serpukhov \cite{AN84} )} \; .\end{array} \right
                       .
\label{poldat1}
\ee
Relaxing the constraint
$\apb = 0$, the Serpukhov data  yield \cite{AN85}
\bea
\apb & = & \;\;1.4 \pm 3.1 \; \mbox{(stat.)} \; \pm 2.5 \; \mbox{(sys.)} \; ,
\nn \\ \amb & = & 15.6 \pm 6.4 \; \mbox{(stat.)} \; \pm 4.4 \; \mbox{(sys.)}
\; ,
\label{poldat2}
\eea
where we have evaluated  $\amb$ from
$\bar{\beta}_{\pi}$ and $\apb$ as given in Ref. \cite{AN85},
adding the errors in quadrature.

 The amplitude $\ggpp$ at low energies is mainly
sensitive to $S$--wave scattering.
In Ref. \cite{KS93} unitarized S--wave
amplitudes have been constructed which contain $\amb$ as an adjustable
parameter. A fit to Mark II data \cite{mark} gives
\bea
\amb & = & \; 4.8 \pm 1.0 \; .
\label{poldat3}
\eea
The result
(\ref{poldat3}) contradicts the Serpukhov analysis Eq. (\ref{poldat1}). Taking
into account also D--waves, Kaloshin et al. \cite{KP94} find
\bea
\apb & = & \left\{ \begin{array}{r@{\: \pm \:} l} 0.22 & 0.06 \;\;\;\; \mbox{(
                       Mark II \cite{mark} )} \\ 0.30 & 0.04 \;\;\; \mbox{(
                       CELLO \cite{cello} ) } \; .  \end{array} \right .
\label{poldat4}
\eea
A detailed analysis of the same data has also been performed in
\cite{MP90}. The authors conclude
 that the
errors quoted in Eq. (\ref{poldat4}) are underestimated,
 see also Ref. \cite{PP94}. By reanalyzing the present data on the 
angular distribution of the $\ggpp$ reaction with using of the unitary model for 
helicity $2$ amplitudes, Kaloshin et al. \cite{KP95} find
\bea
\apb & = & \left\{ \begin{array}{r@{\: \pm \:} l@{\: \pm \:} l} 0.22 & 0.07 \; \mbox{(stat.)} 
                       & 0.04 \; \mbox{(sys.)} \;\;\;\; \mbox{(
                       Mark II \cite{mark} )} \\ 0.33 & 0.06 \; \mbox{(stat.)} & 0.01 \; \mbox{(sys.)} \;\;\;\; \mbox{(
                       CELLO \cite{cello} ) } \; .  \end{array} \right .
\label{poldat5}
\eea
(We have taken into account that the definition of the polarizability in
Refs. \cite{KS93,KP94,KP95} is $4\pi$ larger than the one used here).

Comparing with the data, the two--loop result for $\apb$ agrees within the
error bars with the one found by Kaloshin et al. \cite{KP94,KP95}
(note the remark after Eq. (\ref{poldat4})) and also with the data
from Serpukhov \cite{AN85}. The value for $\amb$ agrees within $1\frac{1}{2}$ 
standard deviation with the measurement 
done at Lebedev \cite{AI86} and also with the data analysis
using unitarized S--waves in \cite{KS93}, however, appears to be inconsistent
with the value quoted in \cite{AN84}.

For a more detailed discussion of the issues considered in this section, we refer the
reader to Ref. \cite{BU96c}. In the following we shortly comment on the relation between the
polarizabilities and the crossed channel reaction $\ggpp$.

\begin{figure}
\unitlength1cm
\begin{picture}(0.5,1) \end{picture}
\epsfysize=9cm \epsffile{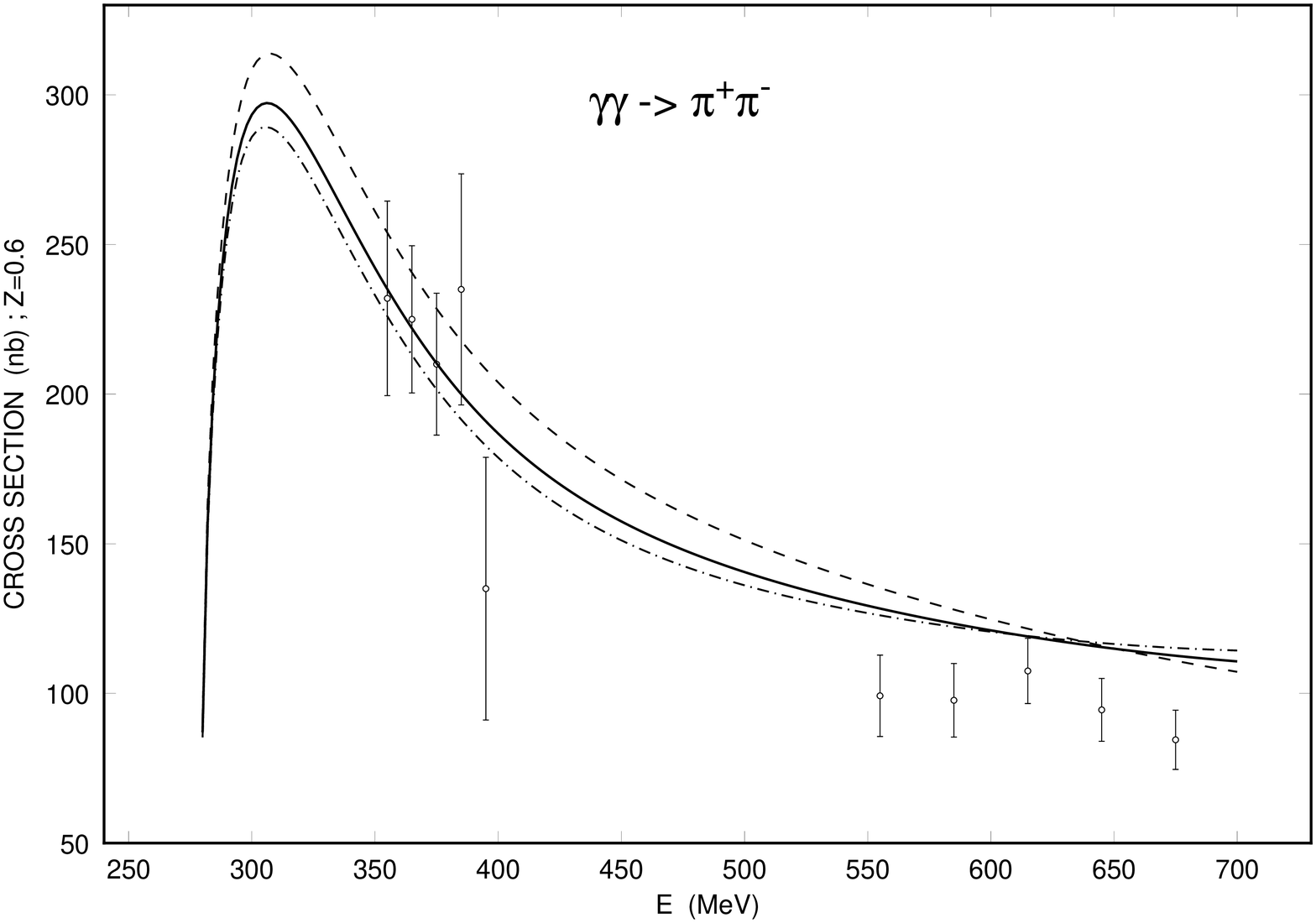}
\caption[Cross section $\ggpp$ parametrized through the
polarizabilities]{\label{ggpppolar}\small{The two--loop result for the $\ggpp$
cross section parametrized through the polarizabilities. We set
$(\alpha+\beta)^{\sss{C}}=0.31$ and $h_s^{r,c}=-0.15$, varying
$(\alpha-\beta)^{\sss{C}}$ between $(\alpha-\beta)^{\sss{C}}=2.21$
(dash--dotted line) and $(\alpha-\beta)^{\sss{C}}=8.84$ (dashed line). The
solid line corresponds to $(\alpha-\beta)^{\sss{C}}=4.42$.}}  \hfill \\[0.2cm]
\begin{picture}(0,1) \end{picture}
\epsfysize=9cm \epsffile{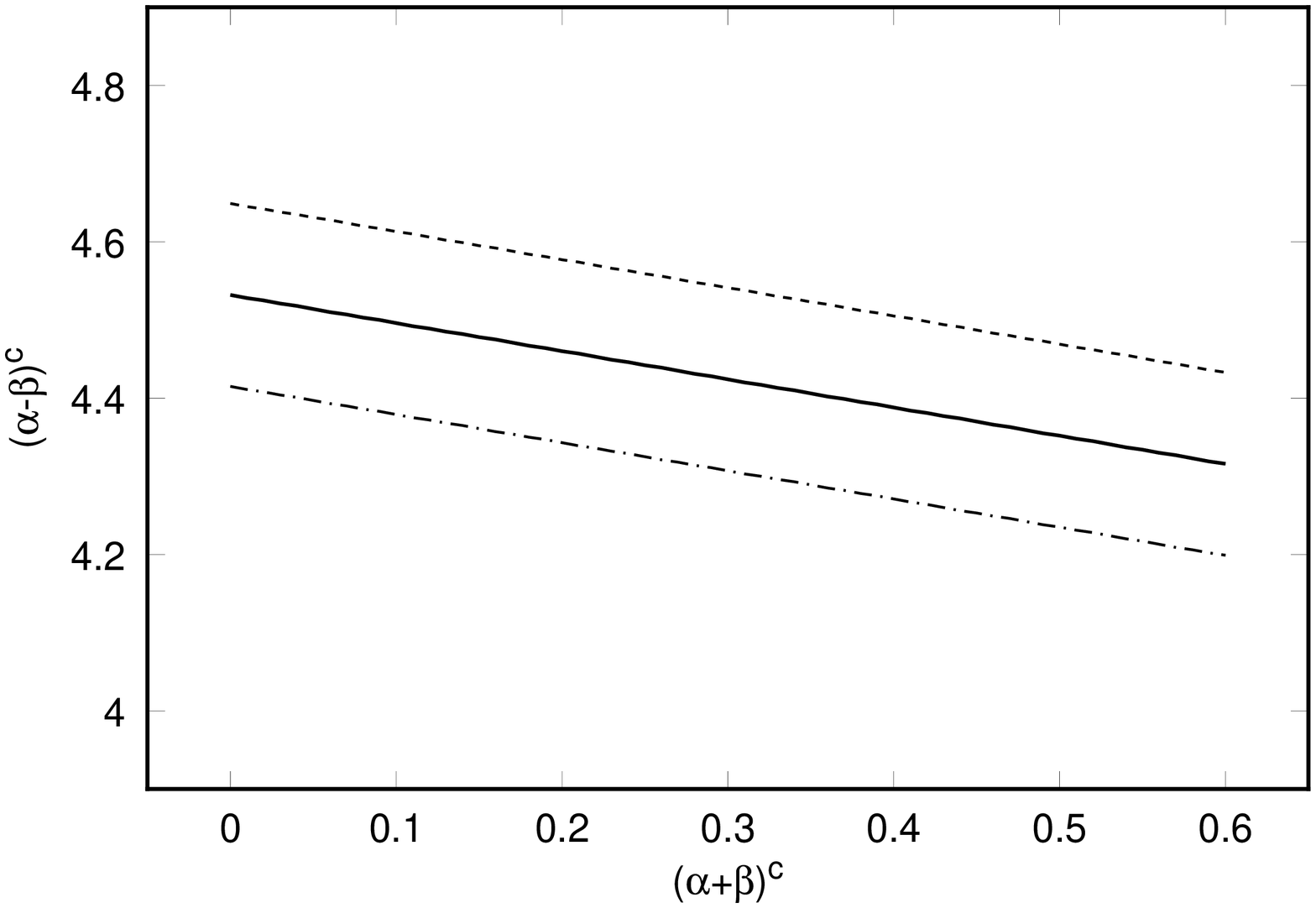}
\caption[Polarizabilities from $\ggpp$ data]{\label{albe}\small{Relation
between $\apb$ and $\amb$ by fixing a point in the $\sigma$--E--plane (E=346
MeV, $\sigma$=223.42 nb), and for different values of $h^{r,c}_s$. The solid
line corresponds to the central value $h^{r,c}_s=-0.15$ whereas the upper
(lower) line is evaluated with $h^{r,c}_s=-0.55 (0.25)$.}}
\end{figure}

\subsection{Pion polarizabilities from $\ggpp$ data}

One may attempt to extract the polarizabilities from $\ggpp$ data in the
following manner \cite{DH93}. Our chiral representation at ${\cal O}(p^6)$ contains three
new low--energy constants $h_{\pm}^{r,c}, h_s^{r,c}$. The two parameters
$h_{\pm}^{r,c}$ may be replaced by the polarizabilities
$(\alpha\mp\beta)^{\sss{C}}$, whereas $h_s^{r,c}$ may be determined e.g. from
resonance exchange. In Fig. \ref{ggpppolar} we have plotted the $\ggpp$ cross
section for a fixed value $(\alpha+\beta)^{\sss{C}}=0.31$ and a fixed value
$h_s^{r,c}=-0.15$, varying $(\alpha-\beta)^{\sss{C}}$ between $2.21$ and
$8.84$. The sensitivity of the cross section to a change in
$(\alpha+\beta)^{\sss{C}}$ is even weaker. Even a $100\%$ change in
$(\alpha\pm\beta)^{\sss{C}}$ is also consistent with the low--energy data. See
also Fig. 9 in Ref. \cite{DH93}.  

On the other hand by choosing a value for
E and $\sigma$ one may extract for a given value $h_s^{r,c}$ the relation
between $(\alpha-\beta)^{\sss{C}}$ and $(\alpha+\beta)^{\sss{C}}$. This is
plotted in Fig. \ref{albe} for different values of $h_s^{r,c}$, where one can
see that the slope of $(\alpha-\beta)^{\sss{C}}$ is small and also changes in
$h_s^{r,c}$ do not affect $(\alpha-\beta)^{\sss{C}}$ significantly. We note
that the sensivity of $(\alpha-\beta)^{\sss{C}}$ due to variation in $\sigma$
or in the low--energy couplings $\bar{l}_i$ is very high. A one percent change
e.g. in the cross section induces a $12\%$ change in
$(\alpha-\beta)^{\sss{C}}$.  

We confirm the statement \cite{DH93} that, using the chiral
representation of the $\ggpp$ amplitude as an interpolation, the statistical
uncertainty of the present low--energy data as well as the uncertainties in the
low--energy couplings do not allow one to pin down the charged pion
polarizabilities to a reasonable accuracy. 
The same is true in the neutral channel \cite{BE94,DH93,JG95}.

\setcounter{equation}{0} \setcounter{subsection}{0}

\section{Summary and conclusions \label{sumsec}}

\begin{itemize}

\item[(i)]

We use $SU(2)\times SU(2) \times U(1)$ chiral perturbation theory to
investigate the $\ggpp$ process up to two loops. We found that far more than 100
different diagrams contribute at ${\cal O}(p^6)$. Among these are
genuine two--loop diagrams that cannot be represented as products of
one--loop contributions (box, vertex, acnode and sunset).
We find it important to take these diagrams into
account, because they contribute substantially even near the threshold. The
three
new low--energy constants appearing at ${\cal O}(p^6)$ have been estimated via
resonance saturation by including vector-- and axial--vector mesons
$(J^{PC}=1^{--},1^{+-},1^{++})$.

\item[(ii)]

The one--loop approximation of the amplitude reproduces the available data
on charged pion pair production from the MARK II experiment
remarkably well. The corrections  at ${\cal O}(p^6)$ induce further small
changes in the cross section, that
then agrees rather well with the dispersion theoretic
calculations \cite{DH93, PE95} at and also substantially above the threshold
region. The enhancement (suppression) of the cross
section very close to threshold (above $E=350 $ MeV) is mainly due to
 $\pi\pi$ rescattering and renormalization of the pion decay
constant,
 see
Fig. \ref{ggpppm1}.
The couplings $h^{r,c}_{\pm}$ and $h^{r,c}_s$ contribute with a negligible
amount below $E=450$ MeV, they become, however, more important at higher
energies.

\item[(iii)]

The cross section for  $\gpmgpm$ is fully dominated by the
Born terms. One-- and two--loop corrections are completely negligible in this
case.

\item[(iv)]

The expansion of the amplitude at the Compton threshold in powers of the 
photon energies allows one to extract the pion polarizabilities in powers of the quark
 masses. 
We note that the sum of the polarizabilities is positive, as is suggested 
by the unsubtracted forward dispersion relation (see e.g. \cite{BU96a}). Furthermore, 
$\apb$ vanishes in the chiral limit $m_u=m_d=0$.
The chiral expansion of the polarizabilities contain chiral logarithms $\sim M_{\pi} \ln^2 M_{\pi}$ and  
$\sim M_{\pi}\ln M_{\pi}$ which are suppressed by small Clebsch--Gordan coefficients.  
Nevertheless their effect on the pion polarizabilities is substantial.
The effect of the low--energy constants $h^{r,c}_{+}$ on the pion polarizability is 
large ($\sim 50 \%$ of the two--loop result). It is therefore essential to make additional efforts 
in estimating the ${\cal O}(p^6)$ low--energy constants in order to reduce the 
theoretical uncertainties in the polarizabilities.

\item[(v)]

We comment on the comparison of the chiral predictions for $\apmb$ with the data.

\begin{itemize}

\item[--]

Our result $\amb = 4.4 \pm 1.0$ includes the leading and next--to--leading order terms. 
It agrees within $1\frac{1}{2}$ standard deviation with the result $40 \pm 24$ found 
at Lebedev \cite{AI86}. On the other hand, it is inconsistent with the value $13.6 \pm 2.8$ 
determined at Serpukhov \cite{AN84}. 

\item[--]

The analysis done by Kaloshin et al. \cite{KS93,KP94,KP95} for $\apmb$ 
agrees within the error bars with the chiral predictions. Beware, however, the remark after 
Eq. (\ref{poldat4}).

\item[--]

The value $\apb = 0.3 \pm 0.1$ includes the leading order term, generated by two--loop graphs. 
It is in good agreement with the result $\apb=0.39 \pm 0.04$ 
\cite{PE81}, obtained from a forward angle dispersion sum rule based on the 
optical theorem and evaluated in a model dependent way. (Note, e.g., that the 
lagrangian used in this model is not chiral invariant.)

\end{itemize}

\item[(vi)]

In principle, the chiral representation of the $\ggpp$ amplitude may
be used as an interpolation to extract the pion polarizabilities from 
the low energy data. However, the cross section is rather insensitive to
$\apmb$, and a reliable determination seems not to be possible in this manner
\cite{DH93,PE95}. 

\item[(vii)]

In order to clarify
the experimental situation,  new experiments to determine the pion
polarizabilities have been planned at Fermilab (E781 SELEX), Frascati (DA$\Phi$NE),
Grenoble (Graal facility) and at Mainz (MAMI). We
refer the reader to the section on hadron polarizabilities in Ref. \cite{BH95}
for details. 

\end{itemize}

\vspace{0.5cm} \noindent {\bf Acknowledgments} \\ [0.3cm] I thank J\"urg
Gasser for his help and advice throughout this work and for reading this
manuscript carefully, Mikko Sainio for useful discussions about numerical
issues and finally Hagen Eck and Sepp Kueblbeck for the update of {\it
FeynArts} to handle CHPT, too.

\newcounter{zahler} \renewcommand{\thesection}{\Alph{zahler}}
\renewcommand{\theequation}{\Alph{zahler}.\arabic{equation}}

\setcounter{zahler}{0}

\appendix

\setcounter{equation}{0} \addtocounter{zahler}{1}

\section{One--loop integrals \label{loopfunc}}
We use dimensional regularization and set
\be
\omega = \frac{d}{2}-2 \le 0.
\ee
Physical results will be obtained by letting $d\rightarrow 4 \;
(\omega\rightarrow 0)$.

\subsection{The integrals $F_m(z)$}
Most of the two--loop graphs can be expressed through the one--loop function
$F_m(z)$. We use the following representation :
\bea
 \int\frac{d^d l}{i(2\pi)^d}\;\frac{1}{[z-l^2]^m} & = & F_m(z) \; .
\label{fm1}
\eea
The $F_m$ satisfy the recursion relation
\be
z\;F_m(z) = \frac{m-3-\omega}{m-1} F_{m-1}(z) \; \;\; , \;\;\; m \ge 2 \; ,
\label{fm2}
\ee
and are given by
\be
F_m(z) = (4\pi)^{-2-\omega}  \frac{\Gamma(m-2-\omega)}{\Gamma(m)} z^{\omega+2-m} \;\;\;
, \;\;\; m \ge 1\; .
\label{fm3}
\ee

\subsection{The loop integrals $J(s)$, $\bar{J}(s)$ and $\stackrel{=}{J}(s)$}

We denote by $J(s)$ the standard one--loop integral $(s \doteq p^2)$
\bea
J(s) & = &\int\frac{d^d l}{i(2\pi)^d}\;\frac{1}{M^2-l^2}\frac{1}{M^2-(l+p)^2} \; .
\label{J1}
\eea
After Feynman--parametrization one gets
\bea
J(s) & = & \int_0^1 \! dx F_2(z) \; , \nn \\ 
z & = & M^2-s x (1-x) \; .
\label{J2}
\eea
$J(s)$ is real analytic in the complex $s$--plane, cut along the positive real
axis for ${\cal R}{\it e} s>4M^2$. The pole at $\omega=0$ may be isolated by writing
\be
J(s) = J(0)+\bar{J}(s) \; ,
\label{J4}
\ee
where $J(0)$ contains the pole,
\be
J(0) = F_2(M^2) = -\frac{1}{16\pi^2}\frac{1}{\omega}+{\cal O}(1) \;\; ; \;\;
\omega\rightarrow 0\; ,
\label{J5}
\ee
and $\bar{J}(s)$ is the finite loop integral
\be
\bar{J}(s) = 
-\frac{1}{16\pi^2}\int_0^1 \! dx \ln \left(1-\frac{s}{M^2} x (1-x )\right)
\; .
\label{J6}
\ee
At small $s$, after expanding the integrand one finds
\be
\bar{J}(s) = \frac{1}{16 \pi^2} \sum_{n=1}^{\infty} \left(\frac{s}{M^2}\right)^n
\frac{(n!)^2}{n(2 n+1)!} \; .
\label{J7}
\ee
For $s > 4M^2$ the logarithm generates an imaginary part
\bea
{\cal J}{\it m}\, \bar{J}(s) & = & \frac{\beta(s)}{16\pi} \;\; , \;\; s>4M^2
\; , \nn \\
\beta(s) & = & \sqrt{1-\frac{4 M^2}{s}} \; .
\label{J8}
\eea
Explicitly,
\be
16 \pi^2 \bar{J}(s) = \left \{ \begin{array}{l@{\quad;\quad}l}
\beta(\ln\frac{1-\beta}{1+\beta}+i\pi)+2 & 4M^2 \le s \\
2-2(\frac{4M^2-s}{s})^{\frac{1}{2}}\arctan (\frac{s}{4M^2-s})^{\frac{1}{2}} & 0
\le s \le 4M^2 \\ \beta\ln\frac{\beta-1}{\beta+1}+2 & s \le 0 \; .
\end{array} \right.
\label{J9}
\ee
In the text we also need
\bea
\stackrel{=}{J}(s) & = & \bar{J}(s) - s \bar{J}'(0) \; , \nn \\ s\bar{J}'(0) &
= & \frac{1}{96\pi^2}\frac{s}{M^2} \; .
\label{J10}
\eea

\subsection{The loop integrals $\bar{G}(s), \stackrel{=}{G}(s)$ and $\bar{H}(s)$}

The integral $\bar{G}(s)$ may be obtained from the tensorial integral
\bea
\int\frac{d^d l}{i(2\pi)^d}\;\left[ \frac{(2 l+q_1)^{\mu}(2 l-q_2)^{\nu}}{D_0}+
g^{\mu\nu} \right] \frac{1}{D_1 D_2}  & = &
\left(g^{\mu\nu}-\frac{2}{s}q_2^{\mu}q_1^{\nu}\right) \bar{G}(s) \; , 
\label{G1}
\eea
\bea
D_0 & = & M^2-l^2 \;\;\; ; \;\;\; D_1 = M^2-(l+q_1)^2 \;\;\; ; \;\;\; D_2
= M^2-(l-q_2)^2 \;,  \nn
\eea
where
\bea
\bar{G}(s) & = &
-\frac{1}{16\pi^2}\left\{1+\frac{2M^2}{s}\int_0^1\frac{dx}{x}\ln
\left(1-\frac{s}{M^2}x(1-x)\right) \right\} \; .
\label{G2}
\eea
$\bar{G}$ is analytic in the complex $s$-- plane, cut along the positive real
axis for ${\cal R}{\it e} s\ge 4M^2$.  At small $s$, after expanding the
integrand one finds
\be
\bar{G}(s) = \frac{1}{16 \pi^2} \sum_{n=1}^{\infty} \left(\frac{s}{M^2}\right)^n
\frac{(n!)^2}{(n+1)(2 n+1)!} \; .
\label{G3}
\ee
For $s > 4M^2$ the logarithm generates an imaginary part
\bea
{\cal J}{\it m}\, \bar{G}(s) & = &
\frac{M^2}{8s\pi}\ln\left\{\frac{1+\beta(s)}{1-\beta(s)}\right\} \;\; , \;\;
s>4M^2, \nn \\ \beta(s) & = & \sqrt{1-\frac{4 M^2}{s}} \; .
\label{G4}
\eea
Use of the Spence function $Li_2$, especially of
\bea
Li_2(y)+Li_2(\frac{-y}{1-y}) & = & -\frac{1}{2}\ln^2(1-y) \; , \nn \\ Li_2(y)
& = & -\int_0^y\frac{dx}{x}\ln(1-x) \; ,
\label{G5}
\eea
gives
\be
-16 \pi^2 \bar{G}(s) = \left \{ \begin{array}{l@{\quad;\quad}l}
1+\frac{M^2}{s}\left(\ln\frac{1-\beta}{1+\beta}+i\pi\right)^2 & 4M^2 \le s \\
1-\frac{4M^2}{s}\arctan^2(\frac{s}{4M^2-s})^{\frac{1}{2}} & 0 \le s \le 4M^2
\\ 1+\frac{M^2}{s}\ln^2\frac{\beta-1}{\beta+1} & s \le 0 \; .
\end{array} \right.
\label{G6}
\ee
In the text we also need
\bea
\stackrel{=}{G}(s) & = & \bar{G}(s) - s \bar{G}'(0) \; , \nn \\ s\bar{G}'(0) &
= & \frac{1}{192\pi^2}\frac{s}{M^2} \; .
\label{G7}
\eea
The loop function $\bar{H}(s)$ is defined in terms of $\bar{G}$ and $\bar{J}$,
\be
\bar{H}(s) = (s-10 M^2) \bar{J}(s)+6 M^2 \bar{G}(s) \; .
\label{G8}
\ee

\end{document}